# One-step theory of fcc-bcc martensitic transformation

C. Cayron

CEA, DRT, LITEN, Minatec, 38054 Grenoble Cedex 9, France.

**Abstract**

Martensitic transformation in steels is responsible for their very high strength and has thus been studied for more than one century since the first works of Martens. However, there is not yet simple physical theory. A rigorous classification of the crystallographic subgroupoids (packets) of the KS variants and the continuity between the KS, NW and Pitsch variants are introduced to represent the crystallographic intricacy associated to the martensite transformation. From this analysis, a new simple "one-step" theory based on Pitsch distortion is proposed. The distortion respects the hard sphere packing of the iron atoms and implies the existence of a neutral line along the close packed directions $[110]_\gamma$ // $[111]_\alpha$. Its principal strains are 0%, -5.8% and 15.5%, well below the +12%, +12%, -20% values of the Bain distortion. Martensite variants nucleate by Pitsch distortion in an austenitic matrix continuously deformed by the transformation. The martensite variants grow by the same Pitsch distortion; they are locally in Pitsch orientation, and therefore are gradually oriented inside the deformation field of austenite leading to the continuum of orientations including KS and NW. Many observations reported in literature are now interpreted, differently than with the usual phenomenological theory. The $\{225\}_\gamma$ habit planes are simply low index $\{112\}_\alpha$ facets of the martensite nucleus. The "twins" sometimes observed at the midrib are actually Pitsch variants. Prior plastic deformation of austenite favours the martensite transformation by probable formation of Lomer-Cottrel locks and distortion field that triggers the Pitsch distortion. Formation of butterfly martensite with internal and external defects is also discussed.

## 1 Introduction

Martensite, generally noted α' or here simply α, is a body centred cubic (bcc) or body centred tetragonal (bct) metastable hard phase obtained in steel and other iron alloys by rapid cooling (quenching) of a high temperature face centred cubic (fcc) phase, the austenite, noted $\gamma$ [1]. It has been named after the German metallurgist Adolf Martens (1850–1914) who studied different steels under optical microscope and established a link between the presence of martensite and the hardness of steels. The $\gamma \rightarrow \alpha$ transformation is in general very rapid (speed of sound in the metal) and diffusionless; all the Fe atoms move collectively to transform the fcc lattice of austenite into the bcc or bct lattice of martensite, trapping the carbon atoms that do not have time to diffuse to form carbides. Martensite starts forming during the cooling at a temperature, Ms, and finishes at another temperature, $M_f < M_s$, and contrarily to bainite, it does not grow when the material is maintained at a temperature between $M_s$ and $M_f$. These temperatures depend highly on the composition of the iron alloy, being higher than 400°C for pure steel and decreases with addition of C or Ni. Martensite appears with different morphologies such as lath, lenticular, butterfly or thin plates, and that order is generally respected when Ms decreases [2]-[4]. Martensite has a specific property: a linear scratch at the surface of a polished surface of austenite is deviated in the part transformed into martensite. This characteristic was supposed to result from a shear mechanism and is at the origin of the crystallographic theories of martensite transformation. Indeed this idea is so impregnated in the metallurgy culture that “martensitic” is now equivalent to “displacive” or “shear” transformation [5]. Since other materials such as NiTi shape memory alloys or transformation-toughened ceramics (zirconia for example) also exhibit lath morphologies supposed to be linked to a shear mechanism, the term martensitic transformation has been extended toward these materials. The martensitic transformation gives the steel their high yield and ultimate strengths, often too high, which makes the steel brittle and necessitates a tempering treatment to let carbon diffuse and steel gains in plasticity. After tempering, martensitic steels still keep impressive yield strength, which make them widely used in construction, automotive and nuclear industries for example. This explains why martensitic transformations have been so much studied. However, despite the huge amount of experimental data and theoretical works, only phenomenological approaches exist for the transformations, and no conclusive, simple and physical theory has been proposed up to now.

### 1.1 Bain distortion and the classical theories

The history of martensitic transformation theories in steels or other Fe alloys can be traced backed to 1924, when Edgar Bain proposed in his paper “The nature of martensite” [6], a simple distortion that allows a fcc lattice to be transformed into a bcc lattice: An intermediate tetragonal lattice is constructed from the fcc one by choosing the ½ $[110]_\gamma$, ½ $[\bar{1}10]_\gamma$ and $[001]_\gamma$ directions as new reference frame and by expanding the two first vectors by 12.6%, and reducing the third one by 20.3%, in order to obtain the bcc lattice with appropriate lattice parameters (Fig. 1).

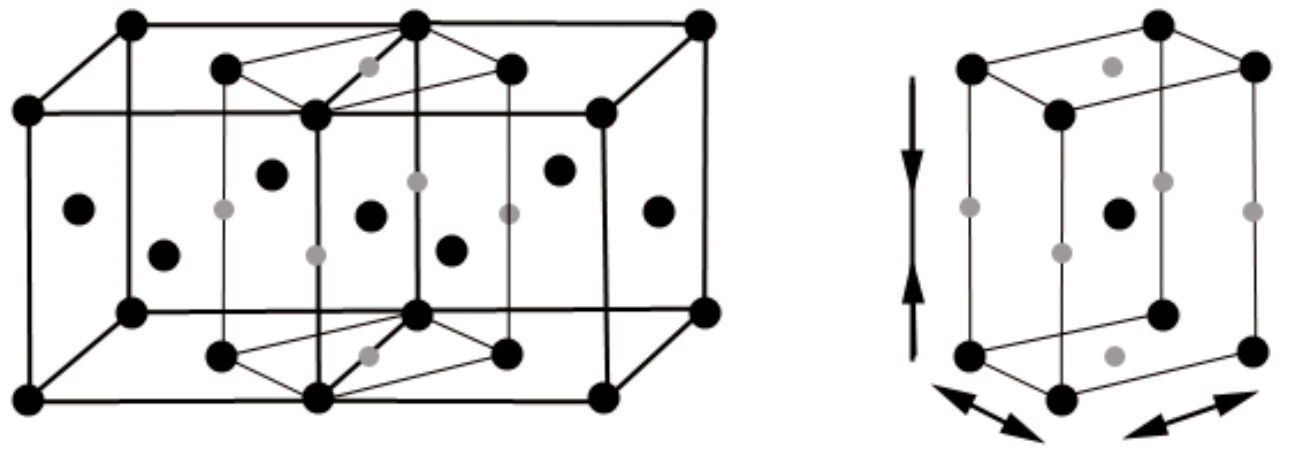

*Fig. 1. Bain distortion (fcc-bct-bcc transformation). The Fe and C atoms are in black and grey, respectively. The distortion is a compression of 20% along the $[001]_\gamma$ axis and expansion of 12% along the $[110]_\gamma$ and $[\bar{1}10]_\gamma$ axes.*

The Bain distortion is often reported to involve the smallest principal strains. We sought the origin and reference of such an affirmation without success, and we will prove later that it is actually false.

The Bain correspondence suffers from intrinsic problems:

- the resulting orientation relationship (OR) between the γ and α phases, called Bain OR, is by more than 10° from the experimental ORs, i.e. Kurdjumov-Sachs [7] and Nishiyama-Wasserman [8][9] ORs measured by X-ray diffraction in 1930's, and the Greninger-Troiano (GT) [10] and Pitsch (P) [11] ORs identified by transmission electron microscopy (TEM) diffraction in the 1950's. More recently a precise average OR was also determined by Miyamoto et *al.* [12] from Electron BackScatter Diffraction (EBSD) measurements. These ORs are given in Table 1.
- as first noticed by Greninger and Troiano [10], the martensite transformation was supposed to result from a shear process, but the shear plane, assumed to be the habit plane, is not in agreement with neither the Bain distortion nor the experimentally observed ORs.

| | |
|---|---|
| Bain : | $(001)_\gamma$ // $(001)_\alpha$ and $[110]_\gamma$ // $[100]_\alpha$ |
| KS: | $(111)_\gamma$ // $(110)_\alpha$ and $[\bar{1}10]_\gamma$ // $[\bar{1}11]_\alpha$ |
| GT: | $(111)_\gamma$ // $(110)_\alpha$ (at 1°) and $[\overline{12},17,\bar{5}]_\gamma$ // $[\overline{17},17,7]_\alpha$ |
| NW: | $(111)_\gamma$ // $(110)_\alpha$ , $[\bar{1}10]_\gamma$ // $[001]_\alpha$ , $[11\bar{2}]_\gamma$ // $[\bar{1}10]_\alpha$ |
| Pitsch: | $(110)_\gamma$ // $(111)_\alpha$ , $[001]_\gamma$ // $[\bar{1}10]_\alpha$ , $[\bar{1}10]_\gamma$ // $[11\bar{2}]_\alpha$ |

*Table 1. The Different orientation relationships observed in martensite: Bain, KS = Kurdjumov-Sachs [7], GT = Greninger-Troiano [10], NW = Nishiyama-Wassermann [8][9], and Pitsch [11]. KS is the most often reported OR in steels and iron alloys. Pitsch and NW are two complementary ORs (the indices $\gamma$ and $\alpha$ are interchanged). Both NW and Pitsch ORs are both at 5° from the KS OR. Pitsch OR has been found in Fe-Ni thin foils of TEM samples. To our knowledge Bain OR has been reported in Fe-Pt alloys [35] but has never been observed in other martensitic iron alloys.*

In order to reconcile the Bain distortion, the measured ORs and HPs, the phenomenological theory of martensite transformation (PTMT), also called phenomenological theory of martensitic crystallography (PTMC), has been developed in the 1950's [13]-[16] (see also [17]-[22] for general review). This theory is in continuity with work of Jawson and Wheeler [23] assuming that the transformation obeys a unique homogeneous strain, but adding now an inhomogenous displacement to council both orientations and shape. It takes the form of sequences of multiplications of matrices, that can be seen as "*reams of indigestible matrix algebra*" [22], each of them representing one part of the problem: a first simple shear $P_1$ (called invariant plane strain IPS) responsible for the macroscopic shape change and habit plane, and a second shear $P_2$ responsible for the structural change (without shape change). This last shear is the superposition of a classical homogeneous deformation and an inhomogenous lattice invariant deformation produced by slip or twinning. The total transformation matrix $T = P_1P_2$ is an invariant line deformation given by the intersection of the two shear plane. The theory assumes that this total transformation matrix can also be written as the initial Bain distortion B associated with a rigid body rotation R so that T = BR. The Bain distortion achieves the desired volume change between the $\gamma$ and $\alpha$ crystals with the "*smallest*" strains. The constant parameter is the Bain matrix, the entry parameters are the lattice constants of the $\gamma$ and $\alpha$ phases and the shear $P_2$, and the exit parameters are the shape shear (which gives the HP) and the OR. There is adjusting parameters, such as the shear and dilatational strains associated to the IPS. The partisans of the PTMT say that it gives the appropriate HP and OR; however there are many hidden or not fully justified assumptions, such as the choice of $P_2$, and more importantly the order of multiplication of matrices whereas the product is non commutative. The PTMT was generalized and further complexified by incorporating multiple shear lattice invariant deformations [24]. Other approaches based on strain energy considerations and interfacial dislocations models are also reported in a recent review on martensitic transformations by Zhang and Kelly [22].

None of the theories mentioned previously is both completely physically supported and self-consistent. There is however in literature an interesting approach that takes its distance with matrix calculation and tries to come back to physics: in 1964, Bogers and Burgers [25] developed an ingenious physical model based on hard sphere representation of the atoms. These researchers noticed that if a shear on a $(111)_\gamma$ plane is stopped at a special position, the operation transforms the 60° angle in two other $\{111\}_\gamma$ planes (angle between the $<110>_\gamma$ directions) into a 70.5° angle of the new $\{110\}_\alpha$ planes (angle between the $<111>_\alpha$ directions). However another shear on another $(111)_\gamma$ plane is required to obtain the final bcc structure. Their work was later refined and promoted by Olson and Cohen [26][27] and can now be summarized as followed: the first shear is on a $\{111\}_\gamma$ plane of vector 1/8 $<112>_\gamma$ direction, which can be achieved by 1/6 $<112>_\gamma$ Shockley partial dislocation averaging one over every second $(111)_\gamma$ slip plane, and the second shear is on another $\{111\}_\gamma$ plane of vector 1/18 $<112>_\gamma$ direction, which can be achieved by 1/6 $<112>_\gamma$ Shockley partial dislocation averaging one over every third $(111)_\gamma$ slip plane. The former is noted T/2 and the latter T/3. This approach is in qualitative agreement with the observations of the martensite formation at the intersection of hcp plates or stacking faulted bands on two $(111)_\gamma$ planes [26][27]. The model has an interesting physical base but its intrinsic asymmetry between one $\{111\}_\gamma$ plane with T/2 and the other $\{111\}_\gamma$ with T/3 seems to be too strict and ideal to be obtained in a real material.

### 1.2 The two-step model and its forgotten ancestors

Recently, we proposed what we believe to be a new mechanism for $\gamma \rightarrow \alpha$ martensitic transformations [28]. Our work came from the observation that in general, there is no one specific OR, but actually all of the ORs reported in Table 1 (except Bain) can be found in the same iron alloy [29][30], with continuum paths between these ORs. These paths take the form of peculiar features in the pole figures (PF) of the martensitic grains forming the prior austenitic grains, which can be reconstructed from EBSD maps with dedicated software [31][32]. We precise here that the continuum paths are not always visible and sufficient spatial resolution should be chosen in the EBSD map to put them in evidence: if the step is too large, only the most representative orientations are acquired and only discrete features representing the average OR are obtained as illustrated in Fig. 2.

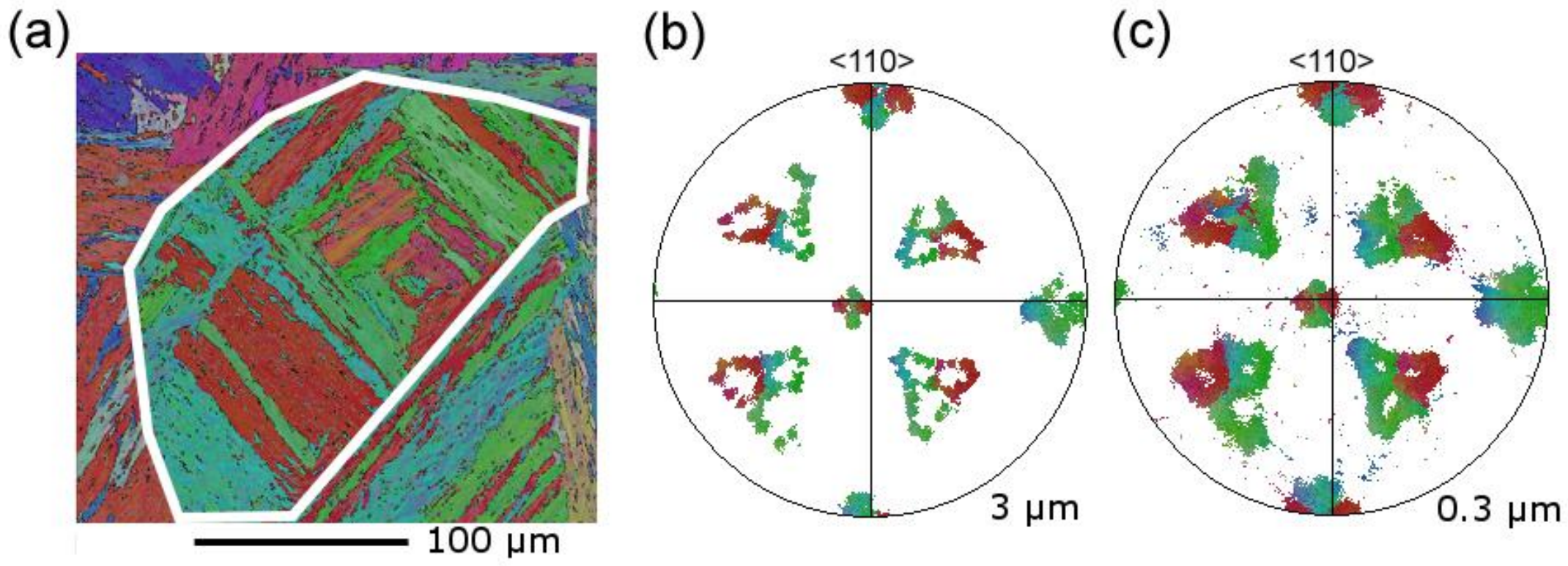

*Fig. 2. Effect of the step size on the pole figure of martensite variants inside a prior austenitic grain. (a) EBSD map of a Fe9CrWTi martensitic steel, with Euler color coding, and <110>$_\alpha$ pole figures of the subset delineated in white, with (b) 3 µm and (c) 0.3 µm for the step size chosen for the EBSD map acquisition.*

The experimental continuous features were simulated by rotating the 24 KS variants with two continuous rotations: the rotations A(*a*) around <111>$_\alpha$ // <110>$_\gamma$ of angle *a* varying between 0 and +10°, and the rotations B(*b*) around <110>$_\alpha$ // <111>γ of angle *b* varying between –6° and 6° or between 0 and 6°. Importantly, we supposed that these two continuous rotations correspond to the trace of the deformation of the fcc matrix imposed by the fcc-bcc transformation. We compared these two rotations to the numerous matrices generated by the PTMT without finding any agreement. Indeed, the Bain distortion or the invariant lattice rotation R should produce continuous features in the PF between the <100>$_\gamma$ direction and the <100>$_\alpha$ directions, and should therefore produce not a Bain ring but a Bain disc or at least a Bain ring with lines inside, which is not the case. The "new" theory we proposed implies the existence of an intermediate hcp phase, called ε: the fcc-bcc transformation would be the result of two steps, a fcc-hcp and hcp-bcc steps which are in agreement with the rotations A and B, respectively. The ε phase is observed in some Fe alloys, such as TRIP steels, but the "two-step" model foresees its existence (even if only fugitive) in all the bcc martensitic steels.

The two-step model is actually very close to the initial model proposed by Kurdjumov and Sachs [7], and by Nishiyama [8] in their original papers, as illustrated in Fig. 3.

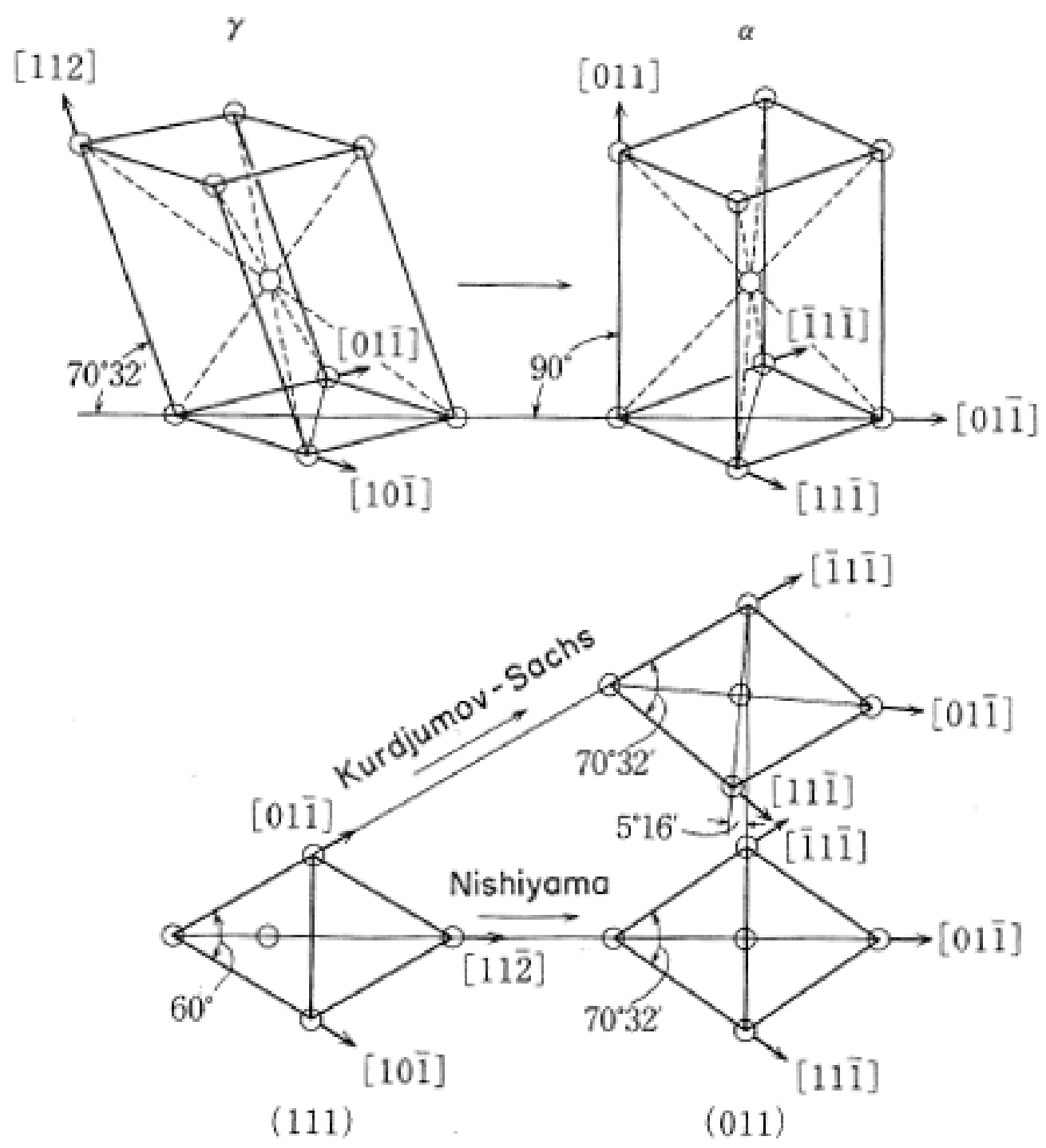

*Fig. 3. KSN model of fcc-bcc transformation by a shear of 19.5° on the (111)γ plane on the [11$\bar{2}$]α direction followed by a distortion of 10.5° (and shuffle). From Nishiyama book [18].*

The KSN model is not anymore presented in the modern books and seems to have been forgotten. In the 1930's, the dislocations were not yet discovered and the partial Shockley dislocations favouring the first step (γ→ε) could not be used as counter-argument to the critics of Greninger and Troiano arguing that « *A more serious objection to these mechanisms is the relatively large movement and readjustments required* » [10] (we wonder why such objection was not also opposed to the Bain distortion at that time). The fact that Nishiyama himself changed his mind and advocated for the PTMT and Bain distortion was probably decisive in the scientific community to make PTMT wins versus KSN[1].

### 1.3 Some limitations of the two-step model

The "new/old" KSN two-step model is simple, based on physical considerations and implies fewer distortions than with Bain correspondence. In order to go deeper in that way, we have realized in-situ ultrafast X-ray diffraction experiments in synchrotron facilities to track the expected fugitive hcp ε phase [33]. The results are under analyzed, but for the moment there is no direct sign of ε phase, even at 6 ms acquisition rate. That is not really surprising because the acquisition rates are still probably not high enough. With a martensite transformation velocity around 1000 m/s, the transformation signal should last less than 1μs. Anyway, there are important obstacles to overcome to build a complete "two-step" theory. Firstly, the rotation A(+10°) is difficult to explain by a shear with partials Shockley dislocations. We tried to find a solution with geometrically necessary dislocations (GND) [34] but we must admit that this approach is not satisfying because the GND formation depends on the many microstructural parameter such as chemistry, grain size etc whereas the continuous features in the PF do not. Moreover the GND should create a rotation field which compensate the shear on the $(111)_\gamma$ plane in the S=$<11\bar{2}>_\gamma$ direction and should rotate the KS variants in the –S direction, whereas we checked afterthought on the PF that the variants are rotated in the +S direction, such that the $[111]_\gamma$ direction is rotated toward the $[110]_\gamma$ direction if the KS OR is respected during the transformation (Fig. 4).

[1] Indeed, in his reference book [18] only three pages are devoted to the KSN model (in the "early shear mechanism models" section) whereas more than hundred pages explain in details the PTMT. The fact that Nishiyama gave up his own theory to adopt the PTMT could be probably related to the development of transmission electron microscopy (TEM). In the 1950's, Nishiyama was one of the first to observe the martensite steels by TEM (firstly on extraction replica) and to identify the numerous stacking fault and nanotwins "predicted" by the PTMT.

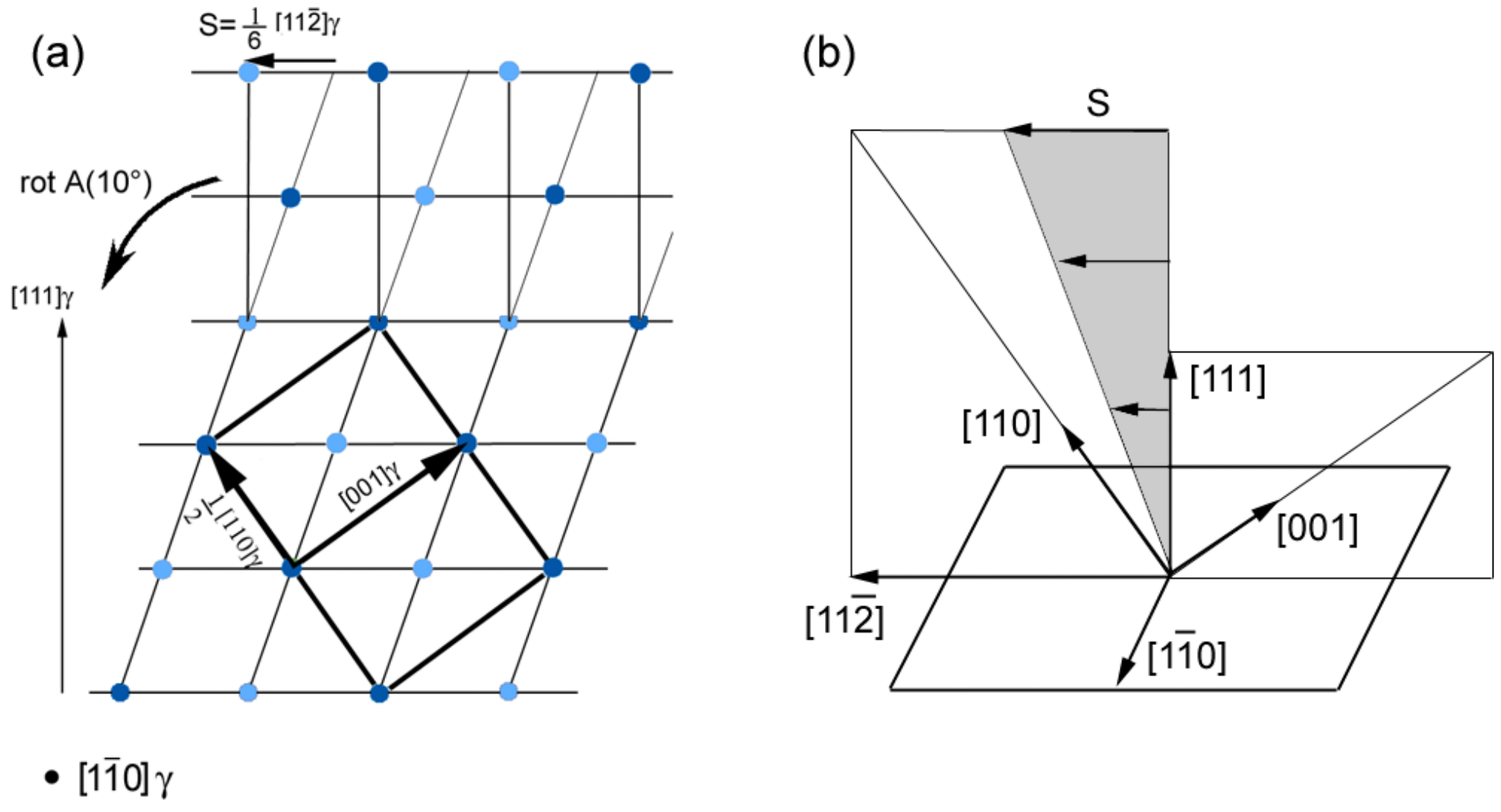

*Fig. 4. Representation of the shear deformation corresponding to fig 5 of [34] with the rotation A(+10°) indicated and (b) equivalent representation with other low indices axes showing that the $[111]_\gamma$ is rotated toward $[110]_\gamma$. This figure is equivalent to Fig. 3 by reversing the direction of x axis.*

Secondly, the large variety of morphologies and habit planes of martensite cannot be explained easily. The first fcc-hcp step should produce a plate on the $(111)_\gamma$ planes and the next hcp-bcc step should give other plates or needles inside the first plate. This is what is actually observed in FeMn alloys in which the ε phase is a stable intermediate phase [35]. It could be possible to cope with that problem by assuming that the two steps occur at the same time. In that case rotations A and B should be correlated. We tried to find an explanation for a possible correlation, while keeping the most important idea of our paper [28]: the pole figures of the martensitic grains exhibit unique singular continuous features whatever the composition and heat treatment of the alloys and whatever the habit planes of martensite and we consider them as the trace of the mechanisms. That was the start of our present research.

The aim of the paper is to present a completely new physical "one-step" model, in which the two previous A and B rotations result from a unique mechanism. We found a distortion tensor with principal strain values lower than with Bain distortion. The model also explains the {225} habit plane, the midribs and other features observed in the past. We already admit that the intermediate hcp ε phase foreseen in our paper [28] has not survived to this new approach. The first section is devoted to the crystallographic/algebraic description of the 24 KS variants and their subgroups. Some results will be used to appropriately describe some configurations of variants in the next sections, but a reader in hurry can skip it at first lecture. The second section is devoted to the model itself. The last section is the interpretation with the new model of some TEM and EBSD observations reported in literature.

# 2 Crystallographic and algebraic study of some configurations of KS variants

## 2.1 The groupoid of KS variants

This section constitutes a crystallographic study of the KS variants and their configurations. The reader not familiar with groups or their extension to groupoids should not be afraid by the vocabulary and understand the terms "groups" or "groupoids" as "sets" or "packets". A groupoid can be simply explained by geometry; it is just a set of objects (here the variants) linked by arrows (the misorientations). The sets of equivalent arrows (i.e. arrows between pairs of variants equivalently misoriented) are called operators. The groupoid composition is a very intuitive law: an arrow from a variant *i* to a variant *j* can be composed with an arrow from variant *j* to variant *k* and the result is the arrow from the variant *i* to the variant *k*, which can be written $[\alpha_i \rightarrow \alpha_j]\ [\alpha_j \rightarrow \alpha_k] = [\alpha_i \rightarrow \alpha_k]$. The arrows from variant *i* to variant *j* can be inverted $[\alpha_i \rightarrow \alpha_j]^{-1} = [\alpha_j \rightarrow \alpha_i]$, and each variant *i* has is own neutral element: the circular arrow $[\alpha_i \rightarrow \alpha_i]$. The groupoids are more generalised than group and are the best mathematical tool to describe a geometrical figure with a global symmetry and composed locally of symmetrical objects. More difficult for non mathematician is probably to translate those notions into algebraic equations: that work implies decomposition of groups into cosets, double-cosets, and to be familiar with group actions. An interested reader can refer to our paper [36] for more details. We will just give the result of such an algebraic approach in the case of the fcc-bcc martensitic transformation with KS OR. Let $\mathbf{G}^{\gamma}$ and $\mathbf{G}^{\alpha}$, the groups of symmetries of the $\gamma$ and $\alpha$ phases, and T the transformation matrix of the parent $\gamma$ crystal to a $\alpha$ daughter crystal, which encodes the OR. The symmetries common to both the parent and daughter crystals forms a subgroup of $\mathbf{G}^{\gamma}$ given by $\mathbf{H}^{\gamma} = \mathbf{G}^{\gamma} \cap \mathrm{T}\, \mathbf{G}^{\alpha}\, \mathrm{T}^{-1}$. The variants are expressed by cosets of type $\alpha_i = g_i^{\gamma} \mathbf{H}^{\gamma}$ with $g_i^{\gamma} \in \mathbf{G}^{\gamma}$ and encoded by set of matrices. Their number results from Lagrange's formula $\mathrm{N}^{\alpha} = |\mathbf{G}^{\gamma}|/|\mathbf{H}^{\gamma}|$. With the KS OR there is only two common symmetries, the identity and inversion symmetry, such that the order of $\mathbf{H}^{\gamma}$ is two and the number of variants is thus 48/2 = 24. The distinct disorientations between the variants, i.e. the operators, are expressed by double cosets, also encoded by set of matrices. Their number is given by the Burnside formula. There are 23 operators between the 24 KS variants, counting the operator identity and distinguishing the polar operators. A polar operator is expressed by a set of equivalent rotations that is distinct of the set of inverses rotations. Non-polar operators are called ambivalent. The 23 operators of the fcc-bcc martensitic transformation with KS OR were calculated with a dedicated computer program called GenOVa [37]; they are given in Table 2. The set of variants and operators form the groupoid of orientational variants [36]. The whole information of this structure is encoded in the groupoid composition table. This table, calculated by GenoVa, is presented in Fig. 5.

It can be read as follow: the first line is composed of operators $\mathrm{O}_n$ with the variants $\alpha_j$ such that the arrow $[\alpha_1 \rightarrow \alpha_j]$ belongs to $\mathrm{O}_n$. The first column is composed of the inverse of operators $\mathrm{O}_m$ with the variants $\alpha_j$ such that $[\alpha_i \rightarrow \alpha_1]$ belongs to the $(\mathrm{O}_m)^{-1}$. The composition of operators is obtained without matrix calculation, directly by the groupoid law: $(\mathrm{O}_m)^{-1}\mathrm{O}_n$ are all the operators that contain the arrows $[\alpha_i \rightarrow \alpha_1]\ [\alpha_1 \rightarrow \alpha_j] = [\alpha_i \rightarrow \alpha_j]$. The product of operators is generally multivalued (see for example the groupoid composition table of NW variants [31]), but with KS OR the product is simply monovalued as a classical mathematical application is. For example $\mathrm{O}_9^{-1}\mathrm{O}_2 = [\alpha_{11} \rightarrow \alpha_1]\ [\alpha_1 \rightarrow \alpha_3] = [\alpha_{11} \rightarrow \alpha_3] \in$ and $= \mathrm{O}_{13}$ (see Table 2).

| | | 0 | 1 | 2 | 3 | 4 | 5 | 6 | 7 | 8 | 9 | 10 | 11 | 12 | 13 | 14 | 15 | 16 | 17 | 18 | 19 | 20 | 21 | 22 | 23 |
|---|---|---|---|---|---|---|---|---|---|---|---|---|---|---|---|---|---|---|---|---|---|---|---|---|---|
| | | 1 | 5 | 3 | 4 | 2 | 6 | 16 | 8 | 9 | 11 | 10 | 17 | 18 | 15 | 14 | 7 | 12 | 13 | 19 | 24 | 21 | 22 | 23 | 20 |
| 0 | 1 | 0 | 1 | 2 | 3 | 4 | 5 | 6 | 7 | 8 | 9 | 10 | 11 | 12 | 13 | 14 | 15 | 16 | 17 | 18 | 19 | 20 | 21 | 22 | 23 |
| 1 | 5 | 4 | 0 | 7 | 10 | 1 | 2 | 3 | 5 | 13 | 18 | 6 | 8 | 9 | 11 | 22 | 17 | 23 | 20 | 12 | 14 | 15 | 16 | 19 | 21 |
| 2 | 3 | 2 | 7 | 0 | 8 | 5 | 4 | 13 | 1 | 3 | 14 | 11 | 10 | 22 | 6 | 9 | 16 | 15 | 21 | 19 | 18 | 23 | 17 | 12 | 20 |
| 3 | 4 | 3 | 9 | 8 | 0 | 15 | 16 | 17 | 14 | 2 | 1 | 12 | 22 | 10 | 21 | 7 | 4 | 5 | 6 | 20 | 23 | 18 | 13 | 11 | 19 |
| 4 | 2 | 1 | 4 | 5 | 6 | 0 | 7 | 10 | 2 | 11 | 12 | 3 | 13 | 18 | 8 | 19 | 20 | 21 | 15 | 9 | 22 | 17 | 23 | 14 | 16 |
| 5 | 6 | 5 | 2 | 1 | 11 | 7 | 0 | 8 | 4 | 6 | 19 | 13 | 3 | 14 | 10 | 12 | 21 | 20 | 23 | 22 | 9 | 16 | 15 | 18 | 17 |
| 6 | 16 | 15 | 3 | 14 | 12 | 9 | 8 | 0 | 16 | 21 | 20 | 17 | 2 | 1 | 22 | 11 | 6 | 19 | 18 | 10 | 7 | 4 | 5 | 23 | 13 |
| 7 | 8 | 7 | 5 | 4 | 13 | 2 | 1 | 11 | 0 | 10 | 22 | 8 | 6 | 19 | 3 | 18 | 23 | 17 | 16 | 14 | 12 | 21 | 20 | 9 | 15 |
| 8 | 9 | 8 | 14 | 3 | 2 | 16 | 15 | 21 | 9 | 0 | 7 | 22 | 12 | 11 | 17 | 1 | 5 | 4 | 13 | 23 | 20 | 19 | 6 | 10 | 18 |
| 9 | 11 | 10 | 18 | 13 | 4 | 17 | 23 | 20 | 22 | 7 | 0 | 9 | 19 | 6 | 16 | 5 | 1 | 2 | 3 | 15 | 21 | 12 | 11 | 8 | 14 |
| 10 | 10 | 9 | 15 | 16 | 17 | 3 | 14 | 12 | 8 | 22 | 10 | 0 | 21 | 20 | 2 | 23 | 18 | 13 | 4 | 1 | 11 | 6 | 19 | 7 | 5 |
| 11 | 17 | 16 | 8 | 9 | 22 | 14 | 3 | 2 | 15 | 17 | 23 | 21 | 0 | 7 | 12 | 10 | 13 | 18 | 19 | 11 | 1 | 5 | 4 | 20 | 6 |
| 12 | 18 | 17 | 10 | 22 | 9 | 18 | 13 | 4 | 23 | 16 | 15 | 20 | 7 | 0 | 19 | 8 | 3 | 14 | 12 | 6 | 5 | 1 | 2 | 21 | 11 |
| 13 | 15 | 14 | 16 | 15 | 21 | 8 | 9 | 22 | 3 | 12 | 11 | 2 | 17 | 23 | 0 | 20 | 19 | 6 | 5 | 7 | 10 | 13 | 18 | 1 | 4 |
| 14 | 14 | 13 | 22 | 10 | 7 | 23 | 17 | 16 | 18 | 4 | 5 | 19 | 9 | 8 | 20 | 0 | 2 | 1 | 11 | 21 | 15 | 14 | 3 | 6 | 12 |
| 15 | 7 | 6 | 12 | 11 | 1 | 20 | 21 | 15 | 19 | 5 | 4 | 18 | 14 | 3 | 23 | 2 | 0 | 7 | 10 | 17 | 16 | 9 | 8 | 13 | 22 |
| 16 | 12 | 11 | 19 | 6 | 5 | 21 | 20 | 23 | 12 | 1 | 2 | 14 | 18 | 13 | 15 | 4 | 7 | 0 | 8 | 16 | 17 | 22 | 10 | 3 | 9 |
| 17 | 13 | 12 | 20 | 21 | 15 | 6 | 19 | 18 | 11 | 14 | 3 | 1 | 23 | 17 | 5 | 16 | 9 | 8 | 0 | 4 | 13 | 10 | 22 | 2 | 7 |
| 18 | 19 | 18 | 17 | 23 | 20 | 10 | 22 | 9 | 13 | 19 | 6 | 4 | 16 | 15 | 7 | 21 | 12 | 11 | 1 | 0 | 8 | 3 | 14 | 5 | 2 |
| 19 | 24 | 23 | 13 | 18 | 19 | 22 | 10 | 7 | 17 | 20 | 21 | 16 | 4 | 5 | 9 | 6 | 11 | 12 | 14 | 8 | 0 | 2 | 1 | 15 | 3 |
| 20 | 21 | 20 | 6 | 19 | 18 | 12 | 11 | 1 | 21 | 23 | 17 | 15 | 5 | 4 | 14 | 13 | 10 | 22 | 9 | 3 | 2 | 0 | 7 | 16 | 8 |
| 21 | 22 | 21 | 11 | 12 | 14 | 19 | 6 | 5 | 20 | 15 | 16 | 23 | 1 | 2 | 18 | 3 | 8 | 9 | 22 | 13 | 4 | 7 | 0 | 17 | 10 |
| 22 | 23 | 22 | 23 | 17 | 16 | 13 | 18 | 19 | 10 | 9 | 8 | 7 | 20 | 21 | 4 | 15 | 14 | 3 | 2 | 5 | 6 | 11 | 12 | 0 | 1 |
| 23 | 20 | 19 | 21 | 20 | 23 | 11 | 12 | 14 | 6 | 18 | 13 | 5 | 15 | 16 | 1 | 17 | 22 | 10 | 7 | 2 | 3 | 8 | 9 | 4 | 0 |

*Fig. 5. KS groupoid table. The first line is composed of operators On (in black) with the variants αj (in blue, below) such that the arrow [α1→αj] belongs to On. The first column is composed of the inverse of operators Om with the variants αi such that [αi→α1] belongs to the (Om)-1. The composition of operators is the operator in box (m,n) given by (Om)-1On = {Op , [αi→α1] [α1→αj] = [αi→αj] ∋ Op }. Here the product is always monovalued, for example O9-1O2 = [α11→α1] [α1→α3] = [α11→α3] ∈ O13 . The KS groupoid can also be represented by a group.*

| | | |
|---|---|---|
| Op. 0 | Id. | [1, 1], [2, 2], [3, 3], [4, 4], [5, 5], [6, 6], [7, 7], [8, 8], [9, 9], [10, 10], [11, 11], [12, 12], [13, 13], [14, 14], [15, 15], [16, 16], [17, 17], [18, 18], [19, 19], [20, 20], [21, 21], [22, 22], [23, 23], [24, 24] |
| Op. 1 | 60.0° / [1 0 1] | [1, 5], [2, 1], [3, 8], [4, 11], [5, 2], [6, 3], [7, 4], [8, 6], [9, 14], [10, 19], [11, 7], [12, 9], [13, 10], [14, 12], [15, 23], [16, 18], [17, 24], [18, 21], [19, 13], [20, 15], [21, 16], [22, 17], [23, 20], [24, 22] |
| Op. 2 | 60.0° / [1 1 1] | [1, 3], [2, 8], [3, 1], [4, 9], [5, 6], [6, 5], [7, 14], [8, 2], [9, 4], [10, 15], [11, 12], [12, 11], [13, 23], [14, 7], [15, 10], [16, 17], [17, 16], [18, 22], [19, 20], [20, 19], [21, 24], [22, 18], [23, 13], [24, 21] |
| Op. 3 | 10.5° / [1 1 1] | [1, 4], [2, 10], [3, 9], [4, 1], [5, 16], [6, 17], [7, 18], [8, 15], [9, 3], [10, 2], [11, 13], [12, 23], [13, 11], [14, 22], [15, 8], [16, 5], [17, 6], [18, 7], [19, 21], [20, 24], [21, 19], [22, 14], [23, 12], [24, 20] |
| Op. 4 | 60.0° / [1 1 0] | [1, 2], [2, 5], [3, 6], [4, 7], [5, 1], [6, 8], [7, 11], [8, 3], [9, 12], [10, 13], [11, 4], [12, 14], [13, 19], [14, 9], [15, 20], [16, 21], [17, 22], [18, 16], [19, 10], [20, 23], [21, 18], [22, 24], [23, 15], [24, 17] |
| Op. 5 | 10.5° / [1 1 0] | [1, 6], [2, 3], [3, 2], [4, 12], [5, 8], [6, 1], [7, 9], [8, 5], [9, 7], [10, 20], [11, 14], [12, 4], [13, 15], [14, 11], [15, 13], [16, 22], [17, 21], [18, 24], [19, 23], [20, 10], [21, 17], [22, 16], [23, 19], [24, 18] |
| Op. 6 | 50.5° / [16 24 15] | [1, 16], [2, 4], [3, 15], [4, 13], [5, 10], [6, 9], [7, 1], [8, 17], [9, 22], [10, 21], [11, 18], [12, 3], [13, 2], [14, 23], [15, 12], [16, 7], [17, 20], [18, 19], [19, 11], [20, 8], [21, 5], [22, 6], [23, 24], [24, 14] |
| Op. 7 | 49.4° / [1 0 1] | [1, 8], [2, 6], [3, 5], [4, 14], [5, 3], [6, 2], [7, 12], [8, 1], [9, 11], [10, 23], [11, 9], [12, 7], [13, 20], [14, 4], [15, 19], [16, 24], [17, 18], [18, 17], [19, 15], [20, 13], [21, 22], [22, 21], [23, 10], [24, 16] |
| Op. 8 | 49.4° / [1 1 1] | [1, 9], [2, 15], [3, 4], [4, 3], [5, 17], [6, 16], [7, 22], [8, 10], [9, 1], [10, 8], [11, 23], [12, 13], [13, 12], [14, 18], [15, 2], [16, 6], [17, 5], [18, 14], [19, 24], [20, 21], [21, 20], [22, 7], [23, 11], [24, 19] |
| Op. 9 | 57.2° / [22 13 26] | [1, 11], [2, 19], [3, 14], [4, 5], [5, 18], [6, 24], [7, 21], [8, 23], [9, 8], [10, 1], [11, 10], [12, 20], [13, 7], [14, 17], [15, 6], [16, 2], [17, 3], [18, 4], [19, 16], [20, 22], [21, 13], [22, 12], [23, 9], [24, 15] |
| Op. 10 | 57.2° / [13 22 26] | [1, 10], [2, 16], [3, 17], [4, 18], [5, 4], [6, 15], [7, 13], [8, 9], [9, 23], [10, 11], [11, 1], [12, 22], [13, 21], [14, 3], [15, 24], [16, 19], [17, 14], [18, 5], [19, 2], [20, 12], [21, 7], [22, 20], [23, 8], [24, 6] |
| Op. 11 | 14.8° / [4 56 21] | [1, 17], [2, 9], [3, 10], [4, 23], [5, 15], [6, 4], [7, 3], [8, 16], [9, 18], [10, 24], [11, 22], [12, 1], [13, 8], [14, 13], [15, 11], [16, 14], [17, 19], [18, 20], [19, 12], [20, 2], [21, 6], [22, 5], [23, 21], [24, 7] |
| Op. 12 | 47.1° / [56 24 49] | [1, 18], [2, 11], [3, 23], [4, 10], [5, 19], [6, 14], [7, 5], [8, 24], [9, 17], [10, 16], [11, 21], [12, 8], [13, 1], [14, 20], [15, 9], [16, 4], [17, 15], [18, 13], [19, 7], [20, 6], [21, 2], [22, 3], [23, 22], [24, 12] |
| Op. 13 | 50.5° / [20 5 16] | [1, 15], [2, 17], [3, 16], [4, 22], [5, 9], [6, 10], [7, 23], [8, 4], [9, 13], [10, 12], [11, 3], [12, 18], [13, 24], [14, 1], [15, 21], [16, 20], [17, 7], [18, 6], [19, 8], [20, 11], [21, 14], [22, 19], [23, 2], [24, 5] |
| Op. 14 | 50.5° / [16 20 5] | [1, 14], [2, 23], [3, 11], [4, 8], [5, 24], [6, 18], [7, 17], [8, 19], [9, 5], [10, 6], [11, 20], [12, 10], [13, 9], [14, 21], [15, 1], [16, 3], [17, 2], [18, 12], [19, 22], [20, 16], [21, 15], [22, 4], [23, 7], [24, 13] |
| Op. 15 | 50.5° / [24 15 16] | [1, 7], [2, 13], [3, 12], [4, 2], [5, 21], [6, 22], [7, 16], [8, 20], [9, 6], [10, 5], [11, 19], [12, 15], [13, 4], [14, 24], [15, 3], [16, 1], [17, 8], [18, 11], [19, 18], [20, 17], [21, 10], [22, 9], [23, 14], [24, 23] |
| Op. 16 | 14.8° / [21 56 4] | [1, 12], [2, 20], [3, 7], [4, 6], [5, 22], [6, 21], [7, 24], [8, 13], [9, 2], [10, 3], [11, 15], [12, 19], [13, 14], [14, 16], [15, 5], [16, 8], [17, 1], [18, 9], [19, 17], [20, 18], [21, 23], [22, 11], [23, 4], [24, 10] |
| Op. 17 | 47.1° / [49 24 56] | [1, 13], [2, 21], [3, 22], [4, 16], [5, 7], [6, 20], [7, 19], [8, 12], [9, 15], [10, 4], [11, 2], [12, 24], [13, 18], [14, 6], [15, 17], [16, 10], [17, 9], [18, 1], [19, 5], [20, 14], [21, 11], [22, 23], [23, 3], [24, 8] |
| Op. 18 | 21.0° / [0 4 9] | [1, 19], [2, 18], [3, 24], [4, 21], [5, 11], [6, 23], [7, 10], [8, 14], [9, 20], [10, 7], [11, 5], [12, 17], [13, 16], [14, 8], [15, 22], [16, 13], [17, 12], [18, 2], [19, 1], [20, 9], [21, 4], [22, 15], [23, 6], [24, 3] |
| Op. 19 | 57.2° / [21 7 18] | [1, 24], [2, 14], [3, 19], [4, 20], [5, 23], [6, 11], [7, 8], [8, 18], [9, 21], [10, 22], [11, 17], [12, 5], [13, 6], [14, 10], [15, 7], [16, 12], [17, 13], [18, 15], [19, 9], [20, 1], [21, 3], [22, 2], [23, 16], [24, 4] |
| Op. 20 | 20.6° / [5 9 9] | [1, 21], [2, 7], [3, 20], [4, 19], [5, 13], [6, 12], [7, 2], [8, 22], [9, 24], [10, 18], [11, 16], [12, 6], [13, 5], [14, 15], [15, 14], [16, 11], [17, 23], [18, 10], [19, 4], [20, 3], [21, 1], [22, 8], [23, 17], [24, 9] |
| Op. 21 | 51.7° / [9 9 5] | [1, 22], [2, 12], [3, 13], [4, 15], [5, 20], [6, 7], [7, 6], [8, 21], [9, 16], [10, 17], [11, 24], [12, 2], [13, 3], [14, 19], [15, 4], [16, 9], [17, 10], [18, 23], [19, 14], [20, 5], [21, 8], [22, 1], [23, 18], [24, 11] |
| Op. 22 | 20.6° / [4 0 13] | [1, 23], [2, 24], [3, 18], [4, 17], [5, 14], [6, 19], [7, 20], [8, 11], [9, 10], [10, 9], [11, 8], [12, 21], [13, 22], [14, 5], [15, 16], [16, 15], [17, 4], [18, 3], [19, 6], [20, 7], [21, 12], [22, 13], [23, 1], [24, 2] |
| Op. 23 | 57.2° / [21 18 7] | [1, 20], [2, 22], [3, 21], [4, 24], [5, 12], [6, 13], [7, 15], [8, 7], [9, 19], [10, 14], [11, 6], [12, 16], [13, 17], [14, 2], [15, 18], [16, 23], [17, 11], [18, 8], [19, 3], [20, 4], [21, 9], [22, 10], [23, 5], [24, 1] |

*Table 2. Operators that link the KS variants, calculated by the software GenOVa [37]. They are given by their index (first column), their angle/axis disorientation (second column) and the set of equivalent arrows, with [i,j] meaning the arrow from the variant αi to the variant αj. The operators O3 and O5 have the lowest disorientation angle (10.5°) around [111]γ and [110]γ, respectively. They are noted in the paper 2A and 2B respectively. Some operators are complementary polar operators such as O9 - O10, Op13 - O14, O19 - O23.*

## 2.2 The subgroupoids of KS variants

For a long time, metallurgists tried to classify some peculiar arrangements or configurations of variants. Gourgues *et al.* [38][39] introduced the notion of morphologic and crystallographic packets, and few later Morito *et al.* [40][41] defined blocks and packets. One must be very careful with those terms, for example the "crystallographic packets" in [38][39] are assembly of variants linked by low misorientations; they constitute therefore what is sometimes called "Bain zones", whereas "packets" are now often used to designate other crystallographic packets, i.e. the sets of 6 variants that share a common $(111)_\gamma = (110)_\alpha$ plane which is also close to the habit plane for low carbon steels [40]. The blocks are pairs of variants interrelated by a $60°/[110]_\alpha$ rotation in a packet and forming parallel laths with similar contrast in optical microscopy [41]. We have represented in the Fig. 6 the 24 KS variants corresponding to the Table 2 and Fig. 5 such that the $\{111\}_\gamma$ faces of the γ parent phase are in white and the $\{110\}_\alpha$ planes in green.

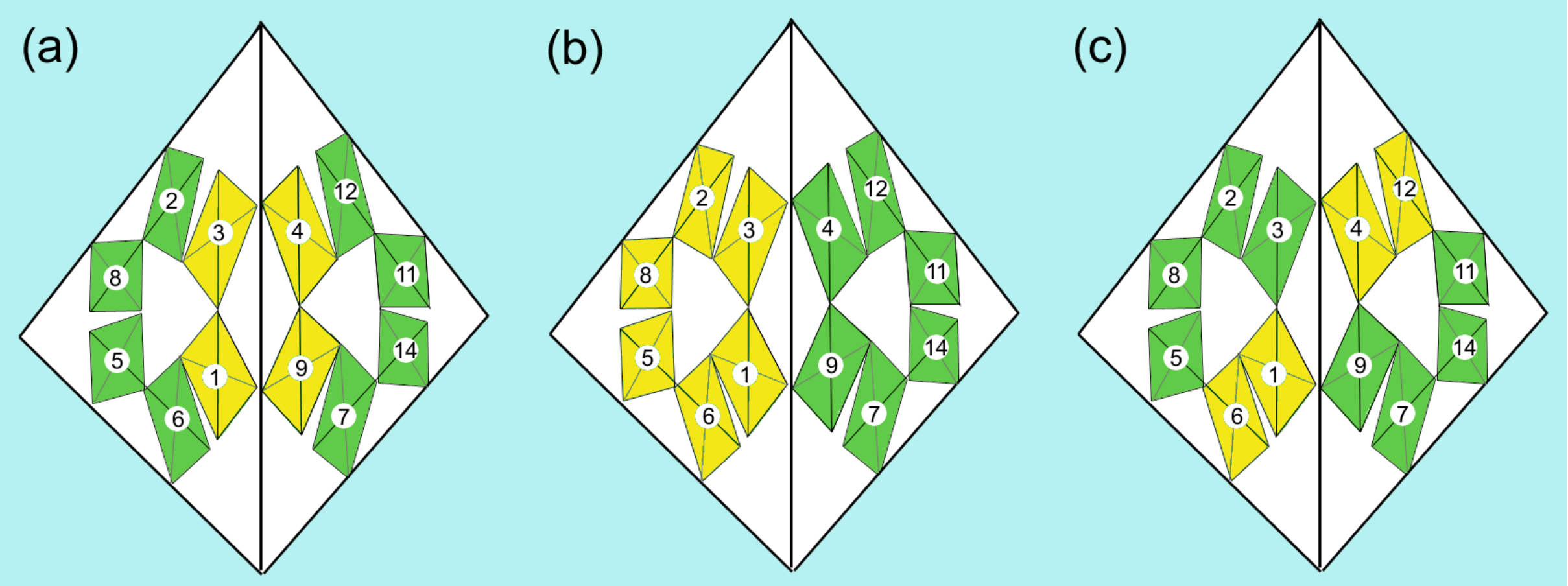

*Fig. 6. Three important crystallographic KS packets represented in 3D and marked in yellow: (a) CPD packet, (b) CPP packet and (d) Bain packet.*

Three special configurations, CPD, CPP and Bain, are marked in yellow and detailed in the following. They result from our classification of subsets of variants only on crystallographic/ algebraic considerations; more explicitly, by considering the subgroupoids of the groupoid of the 24 KS variants with its 23 operators, without any consideration on the morphologies and habit planes. From Table 2, it can be noticed that there are two kinds of low-misoriented pairs of variants: the first pairs linked by the operator $O_3$ = 10.5° / $[111]_\alpha$ that will be called 2A, and the pairs linked by the operator $O_5$ = 10.5° / $[110]_\alpha$ that will be called 2B. This last pairs form the blocks defined by Morito *et al.* [40]. The notation "2A" and "2B" come from the choice we made (and that will be justified later) to name the rotation A(5.25°) = A and B(5.25°) = B. Three crystallographic packets can be built with the operators 2A, 2B and 2A+2B:

<u>CPD packets:</u> Some variants share a common $<111>_\alpha$ axis which is parallel to one $<110>_\gamma$ direction. It is possible to prove with the Table 2 and Fig. 5 that the sets of such variants and their operators form close assemblies, i.e. a subgroupoids. Since $<111>_\alpha$ and $<110>_\gamma$ axes are the dense directions of the respective phase, we will call those subgroupoids the Close Packed Direction (CPD) subgroupoids, or simply CPD packets. There are 6 CPD packets that each contains the operators $O_2$, $O_3$ and $O_8$. The CPD packets can be generated by one variant and two operators $O_2$ and $O_3$. They are explicitly given as sets of variants in Table 3.

| Name | Variants inside the packets | Operators inside the packets |
|---|---|---|
| CPD packets | CPD-1 = $\{\alpha_1, \alpha_3, \alpha_4, \alpha_9\}$<br>CPD-2 = $\{\alpha_2, \alpha_8, \alpha_{10}, \alpha_{15}\}$<br>CPD-3 = $\{\alpha_5, \alpha_6, \alpha_{16}, \alpha_{17}\}$<br>CPD-4 = $\{\alpha_7, \alpha_{14}, \alpha_{18}, \alpha_{22}\}$<br>CPD-5 = $\{\alpha_{11}, \alpha_{12}, \alpha_{13}, \alpha_{23}\}$<br>CPD-6 = $\{\alpha_{19}, \alpha_{20}, \alpha_{21}, \alpha_{24}\}$ | $\{O_0, O_2, O_3$ and $O_8\}$ |
| CPP packets | CPP-1 = $\{\alpha_1, \alpha_2, \alpha_3, \alpha_5, \alpha_6, \alpha_8\}$<br>CPP-2 = $\{\alpha_4, \alpha_7, \alpha_9, \alpha_{11}, \alpha_{12}, \alpha_{14}\}$<br>CPP-3 = $\{\alpha_{10}, \alpha_{13}, \alpha_{15}, \alpha_{19}, \alpha_{20}, \alpha_{23}\}$<br>CPP-4 = $\{\alpha_{16}, \alpha_{17}, \alpha_{18}, \alpha_{21}, \alpha_{22}, \alpha_{24}\}$ | $\{O_0, O_1, O_2, O_4, O_5$ and $O_7\}$ |
| Bain packets | Bain-1 = $\{\alpha_1, \alpha_4, \alpha_6, \alpha_{12}, \alpha_{17}, \alpha_{19}, \alpha_{21}, \alpha_{23}\}$<br>Bain-2 = $\{\alpha_5, \alpha_8, \alpha_{11}, \alpha_{13}, \alpha_{14}, \alpha_{15}, \alpha_{16}, \alpha_{22}\}$<br>Bain-3 = $\{\alpha_2, \alpha_3, \alpha_7, \alpha_9, \alpha_{10}, \alpha_{18}, \alpha_{20}, \alpha_{24}\}$ | $\{O_0, O_3, O_5, O_{11}, O_{16}, O_{18}, O_{20}$ and $O_{22}\}$ |

*Table 3. Some important subgroupoids (packets) of KS variants: close packed direction (CPD), close packed plane (CPP) and Bain packets, with their sets of variants and operators.*

CPP packets: Some variants share a common $\{110\}_\alpha$ plane which is parallel to one $\{111\}_\gamma$ plane. Even if obvious geometrically, here again it can be proved that the sets of such assemblies of variants are close under the groupoid actions and are therefore subgroupoids. Since $\{110\}_\alpha$ and $\{111\}_\gamma$ planes are the dense planes of the respective phase, we will call those subgroupoids the Close Packed Plane (CPP) subgroupoids, or simply CPP packets. There are 4 CPP packets that each contains the operators $O_1$, $O_2$, $O_4$, $O_5$ and $O_7$. The CPP packets can be generated by one variant and two operators $O_1$ and $O_5$. They correspond to the "packets" defined by Morito *et al.* [40]. They are given as sets of variants in Table 3.

Bain packets: Some variants are linked between them by low misorientations. They are the crystallographic packets defined by Gourgues *et al.* [38], also called Bain zones. The two lowest misorientations are the 2A and 2B operators, i.e. $O_3$ = 10.5° / $[111]_\alpha$ and $O_5$ = 10.5° / $[110]_\alpha$. May the structure, generated by one variant and these two operators, be really close? Would it not form the whole groupoid? Since the solution is here not obvious geometrically (at least at first thought) we decide to prove it. Let's consider the variant $\alpha_1$ and build the other variants generated by the operators 2A and 2B. The result is given graphically in Fig. 7. The set of generated variants is close and contains 8 variants. Therefore there are 3 Bain packets. Each packet forms in the $<100>_\alpha$ pole figures of the 24 KS a Bain circle around one of the three $<100>_\gamma$ directions.

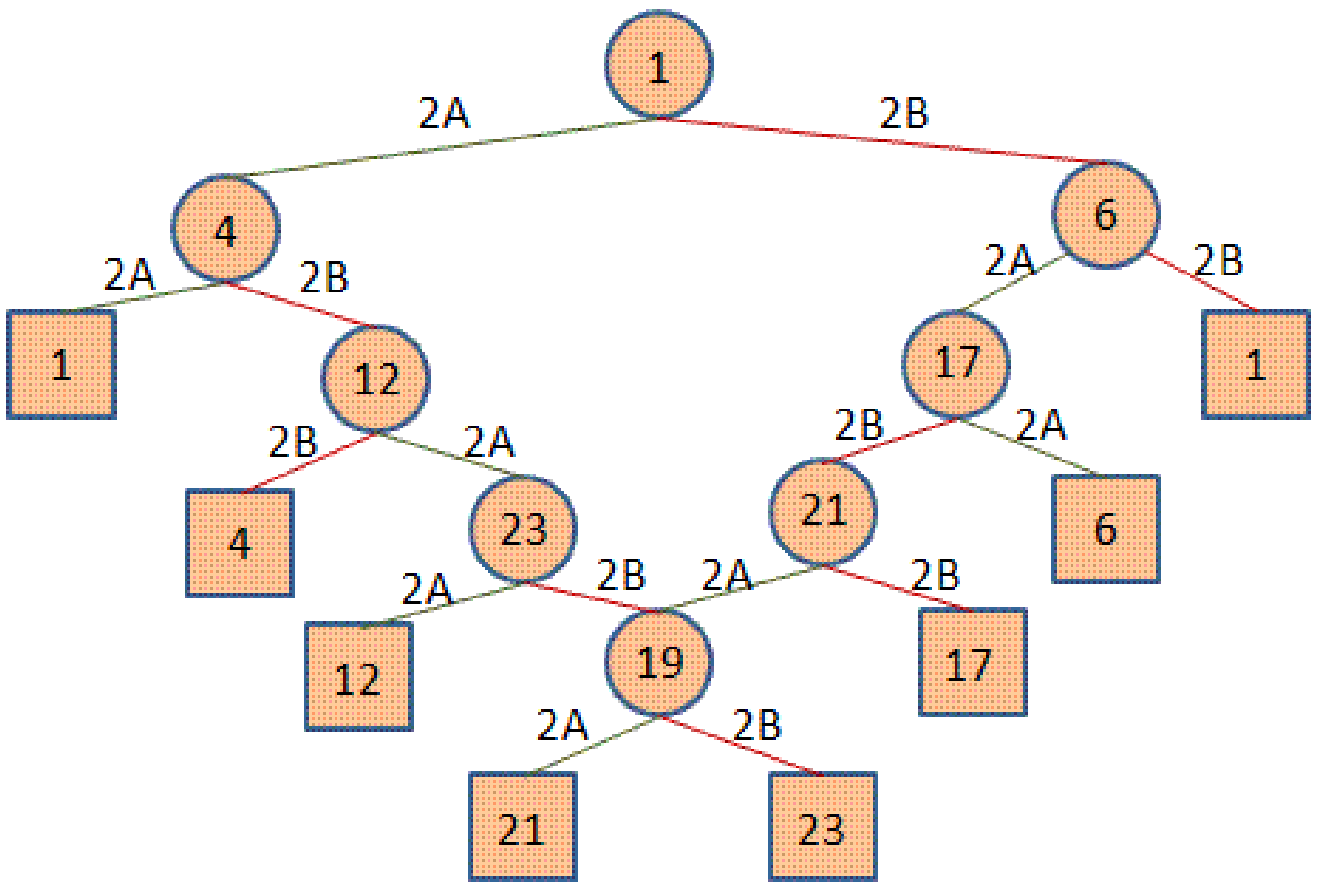

*Fig. 7. A Bain packet in the tree form. This subgroupoid is generated by the variant $\alpha 1$ and the two low-misorientation operators 2A= O3 and 2B = O5. The numbers in the circles are the indexes of the KS variants.*

## 2.3 The closing rotations A and B

When we published in 2010 the two-step model, it was well known that the rotation B = B(5.25°) allowed the continuous path between the KS variants and the NW variants, and in our approach that path was the result of the hcp-bcc distortion by a Burgers mechanism [28]. Both KS or NW ORs appeared as the result of the final step depending on a final preference (NW or KS OR) that could not be explained. The NW variants can be easily imagined on Fig. 6b: they are on the CPP packets at the mid-positions between the pairs linked by the operator $O_3$ = 2B, such as $\alpha_1$-$\alpha_6$, $\alpha_2$-$\alpha_3$, $\alpha_5$-$\alpha_8$. The B path transforms the two KS variants of the pairs into one NW variant, as simulated on the pole figure of Fig. 8a and represented in Fig. 8b. The angular value of 10.5° of $O_3$ results from the distortion of the 60° angle of the $<110>_\gamma$ directions in $(\bar{1}11)_\gamma$ plane into the 70.5° angle of the $<111>_\alpha$ directions in the $(110)_\alpha$ plane. The angle 5.25° of B is the half value of the difference of these two angles and is therefore:

$$ang(B) = \frac{1}{2}(\arccos(1/3) - \arccos(1/2)) = \frac{1}{2}\arccos(\sqrt{2/3} + 1/6) \approx 5.25° \qquad (1)$$

The rotation B acts as a closing operation on the structure of the 24 KS variants, and the 12 NW variants constitute a locking configuration.

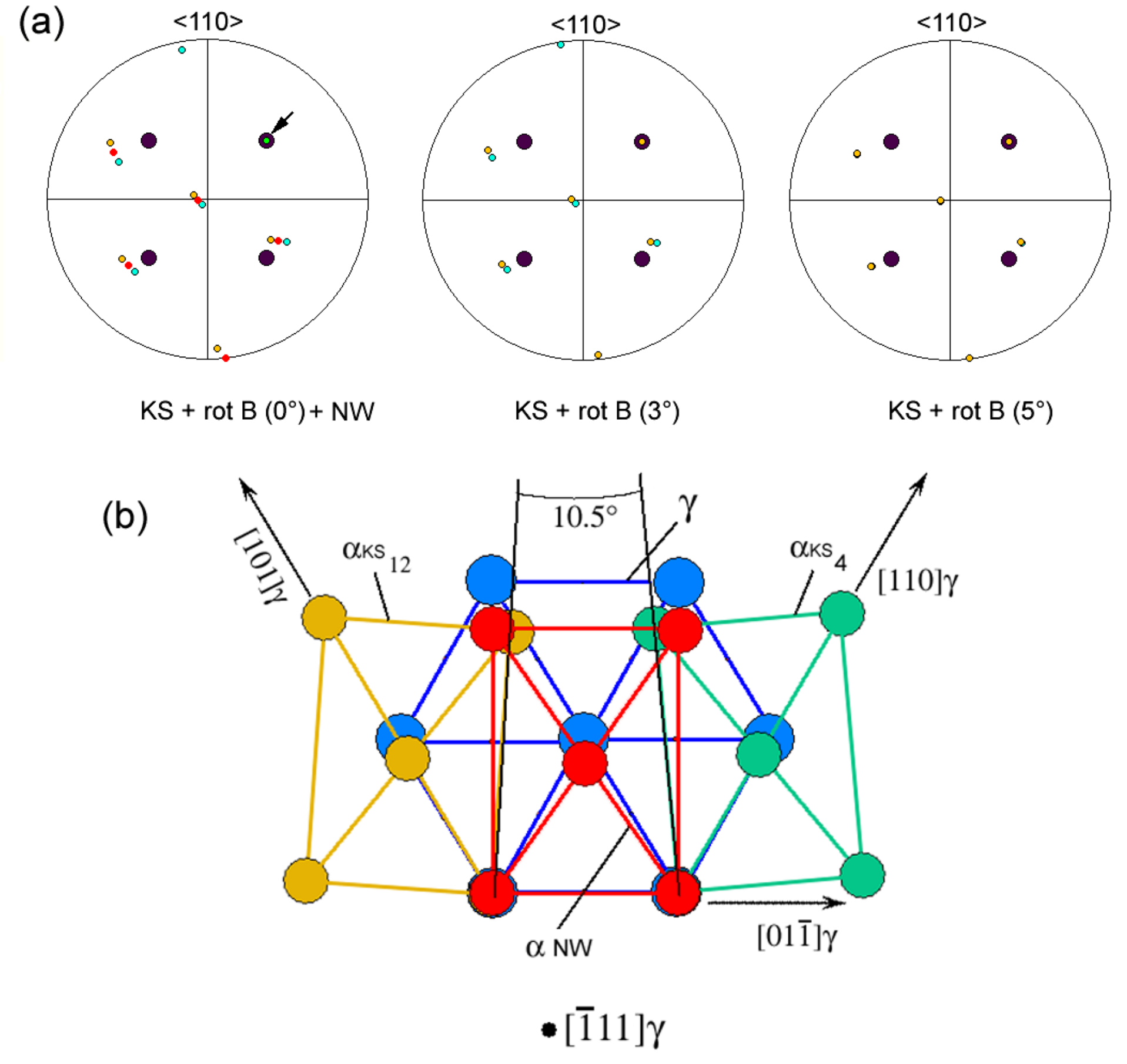

*Fig. 8. Effect of the continuous rotation B($\theta$) on the pair of KS variants $\alpha_4$ - $\alpha_{12}$ (linked by the operator $O_5$) viewed on the <110>$_\alpha$ pole figure, with $\theta$ = 0°, $\theta$ = 3° and $\theta$ = 5°. For this last angle (5.25° more precisely), the two variants overlap into only one NW variant. (b) Explanation on the direct space with $\gamma$ oriented along the $[\bar{1}11]_\gamma$ // $[110]_\alpha$ direction. The KS $\alpha_4$ and $\alpha_{12}$ variants are in green and yellow respectively, and the NW variant in red.*

At that time in 2010 we imagined that the continuous path of A(10.5°) determined from the experimental EBSD pole figures was the consequence of a shear on the $(111)_\gamma$ plane inducing the first fcc-hcp step. We missed something important in our analysis: we did not realized that the rotation A(10.5°) was an operator between two KS variants, i.e. the operator $O_5$, that we named 2A. It links some pairs of variants in the CPD packets, such as $\alpha_1$-$\alpha_4$, $\alpha_3$-$\alpha_9$. One can ask: are there intermediate variants for 2A, as there are for 2B (the NW variants)? If yes, what are they? By simulating the effect of rotation A on the KS variants Fig. 9a we show that the answer is yes. The representation in Fig. 9b helps to understand why. The angular value of 10.5° of the $O_5$ results from the distortion of the 60° angle of the <110>$_\alpha$ directions in the $(111)_\alpha$ plane into the 70.5° angle of the <111>$_\gamma$ directions in the $(110)_\gamma$ plane. The angle 5.25° of A is therefore here again:

$$ang(A) = \frac{1}{2}(\arccos(1/3) - \arccos(1/2)) = \frac{1}{2}\arccos(\sqrt{2/3} + 1/6) \approx 5.25° \qquad (2)$$

The rotation A acts as a closing operation on the structure of KS variants. The intermediate variants are such that the $(111)_\alpha$ plane is parallel to the $(110)_\gamma$ plane and the $[\bar{1}10]_\alpha$ direction is parallel to the $[001]_\gamma$ direction. We realized that these variants are in Pitsch OR. The rotation A acts as a closing operation on the structure of the 24 KS variants, and the 12 Pitsch variants constitute another locking configuration.

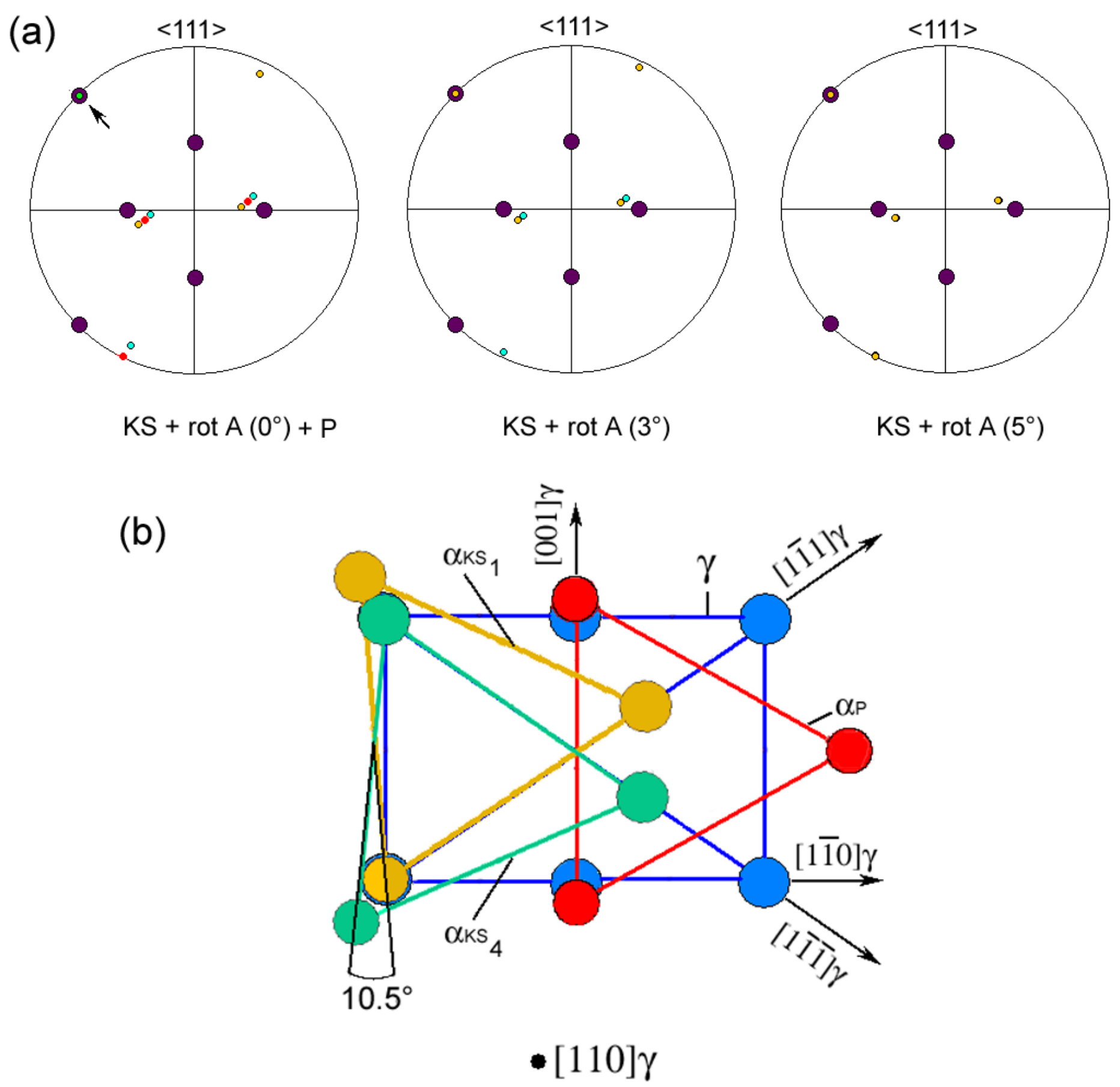

*Fig. 9. Effect of the continuous rotation A(θ) on the pair of KS variants α1 - α4 (linked by the operator O3) viewed on the <111> α pole figure, with θ = 0°, θ = 3° and θ = 5°.   For this last angle (5.25° more precisely), the two variants overlap into only one Pitsch variant. (b) Explanation on the direct space with γ oriented along the [110]γ // [111]α direction. The KS α1 and α4 variants are in green and yellow respectively, and the Pitsch variant in red.*

This analysis shows that the rotations A and B acting on the 24 KS variants create intermediate variants that close the structure: rotation A close the $<111>_{\alpha}$ CP directions via the 12 Pitsch variants, and rotation B close the $\{110\}_{\alpha}$ CP planes via the NW variants. We have simulated the pole figures as we did in our paper [28], but now by limiting the angle *a* of the rotation A to 5.25° (and not 10.5°). As expected, the simulated pole figures, given in Fig. 10, are the same as those presented in our paper [28]. It is also possible to slightly improve the fit between the simulated and experimental pole figures by combining the rotations A and B i.e. by introducing the rotation A(*c*)+B(5.25°-*c*) with $c \in [0,5.25°]$. The part generated by A with *a* between 5.25° and 10.5° in paper [28] was actually overlapping the part generated by A with *a* between 0° and 5.25° and was unnecessary.

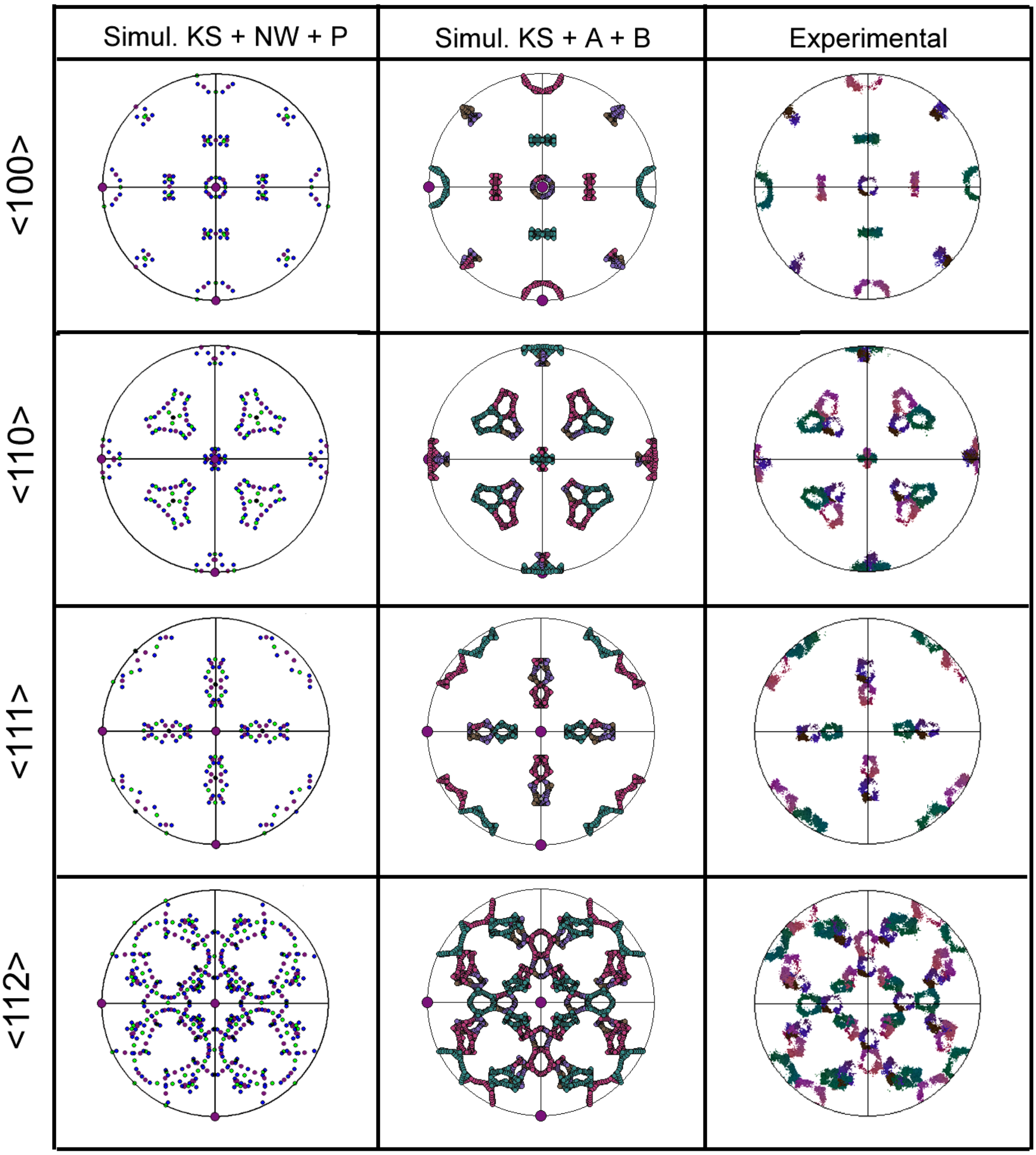

*Fig. 10. Pole figures of martensitic variants. First column: the 24 KS, 12 NW and 12 Pitsch variants in blue, purple and green respectively. Second column, the 24 KS variants with continuous rotations A(a) and B(b) with $a \in [0,5.25°]$ and $b \in [0,5.25°]$, and their composition A + B = A(c)+B(5.25°-c) with $c \in [0,5.25°]$. Third column: experimental pole figures of martensitic variants in a prior austenitic grain of a Fe9Cr (EM10) steel.*

Our approach is very effective to represent the high crystallographic intricacy of the martensite variants. All the KS, NW and Pitch variants are now closed on themselves by rotations A and B and constitute a close structure, like a nutshell. Is it possible to find inside it one physical mechanism that could explain such a fascinating intricacy and at the same time explains the martensitic transformation? In that aim, from which OR should we start our research: KS, NW or Pitsch? Which side must we take to crack the nut?!

# 3 One-step model based on Pitsch distortion

The logical OR to build a model should be NW or KS because they are the two most often reported ORs in steels. However, that way has been explored for more than 60 years and has leaded to (a) the complex and somehow artificial phenomenological theory, (b) the interesting but too strict Bogers and Burgers model, and (c) the two-step KSN model with its limitations (see section 1). Could Pitsch OR be interesting to build a new theory? At first look, the answer seems no. Pitsch reported that OR after an austenitization and rapid cooling of a thin TEM lamella of a iron-nitrogen alloy. Pitsch transposed the OR into a tensor, composed of a pure distortion and half twin shear, without according a fundamental role to it, except a "formal" discussion within the conventional Bain distortion and PTMT. The Pitsch OR and related distortion were considered as an exotic result of a dimensional effect which allows stress relaxation on thin foils or surfaces. Nobody, even Pitsch himself, has investigated the Pitsch distortion in detail, at atomic scale. We did it, and realized that it gives what we believe to be the key to understand fcc-bcc martensitic transformation.

## 3.1 Decomposition of the Pitsch distortion

In the following, we will assume that the $\alpha$ phase results from the $\gamma$ phase such that atoms are hard spheres of same diameter in both phase. By considering the $<110>_\gamma$ and $<111>_\alpha$ dense directions, this assumption implies that:

$$\sqrt{2}\, a_\gamma = \sqrt{3}\, a_\alpha \qquad (3)$$

The lattice parameter of $a_\gamma$ = 0.3585 nm should give $a_\alpha$ = 0.2927 nm, whereas it is actually $a_\alpha$ = 0.2866 nm. The difference of 2% can be attributed to electronic and magnetic properties of iron. It will not be taken into consideration in the following for sake of simplicity.

The Pitsch OR is $[110]_\gamma // [111]_\alpha$ , $[\bar{1}10]_\gamma // [11\bar{2}]_\alpha$ and $[001]_\gamma // [\bar{1}10]_\alpha$. These axes form a orthogonal (but not orthonormal) reference basis $B_1$ = (x,y,z). From the hard sphere assumption, we can precise that along x, the parallelism condition is actually an equality $[110]_\gamma = [111]_\alpha$, which means that x is a neutral (invariant) line. The complete Pitsch distortion is represented in Fig. 11. The initial γ lattice is shown in Fig. 11a such that the $(\bar{1}10)_\gamma$ plane is horizontal and $(110)_\gamma$ plane vertical; the distortion is represented in Fig. 11b (without showing the atoms at centres of the faces of the γ phase in order to make the figure simpler), the resulting α crystal -actually a tetragonal frame of it- is shown in Fig. 11c, and its basic lattice is shown in Fig. 11d. In order to make the figure easier to visualize, the atoms have not their real size, and the atoms at the centres are a little smaller that those at the corners. However, it is important to keep in mind that the hard sphere atoms of the γ phase along the x and y axis in Fig. 11a keep the contact during the transformation along the neutral line x, and also along the y /y' directions (y for γ phase and y' for α phase), as explained in the following:

We consider first the distortion of the $(001)_\gamma$ plane into $(\bar{1}10)_\alpha$ plane, viewed perpendicularly along the z axis (Fig. 12a). The angle of 90° between the $[110]_\gamma$ and $[\bar{1}10]_\gamma$ diagonals of the $(001)_\gamma$ square is distorted into the angle of 70.5° between the $[111]_\alpha$ and $[11\bar{1}]_\alpha$ diagonals of the $(\bar{1}10)_\alpha$ rectangle. In addition the $[110]_\gamma$ = $[111]_\alpha$ direction remains unchanged (no dilatation and no rotation). Therefore the distortion is completely obtained by the rotation of y = $[\bar{1}10]_\gamma$ by $\alpha$ =19.5° around the common $[001]_\gamma$ // $[\bar{1}10]_\alpha$ axis, which becomes after rotation y' = $[11\bar{1}]_\alpha$. This distortion is a pure rotation because these directions are close packed directions. It is represented by a vector δ. As shown in Fig. 12b, the coordinates of δ are $\delta_x = \sin(\alpha) = 1/3$, $\delta_y = \cos(\alpha)-1 = \sqrt{8}/3-1$, $\delta_z = 0$. In rough approximation, if one assume that $\delta_y << \delta_x$ , the distortion δ appears as a shear of value $\delta_x$ = 1/3 in the $[111]_\alpha$ direction on the $(11\bar{2})_\alpha$ plane. We think that Pitsch did such an approximation when he spoke about a half shear of the α phase. However, such approximation is not useful and all the components of δ will be kept in the following.

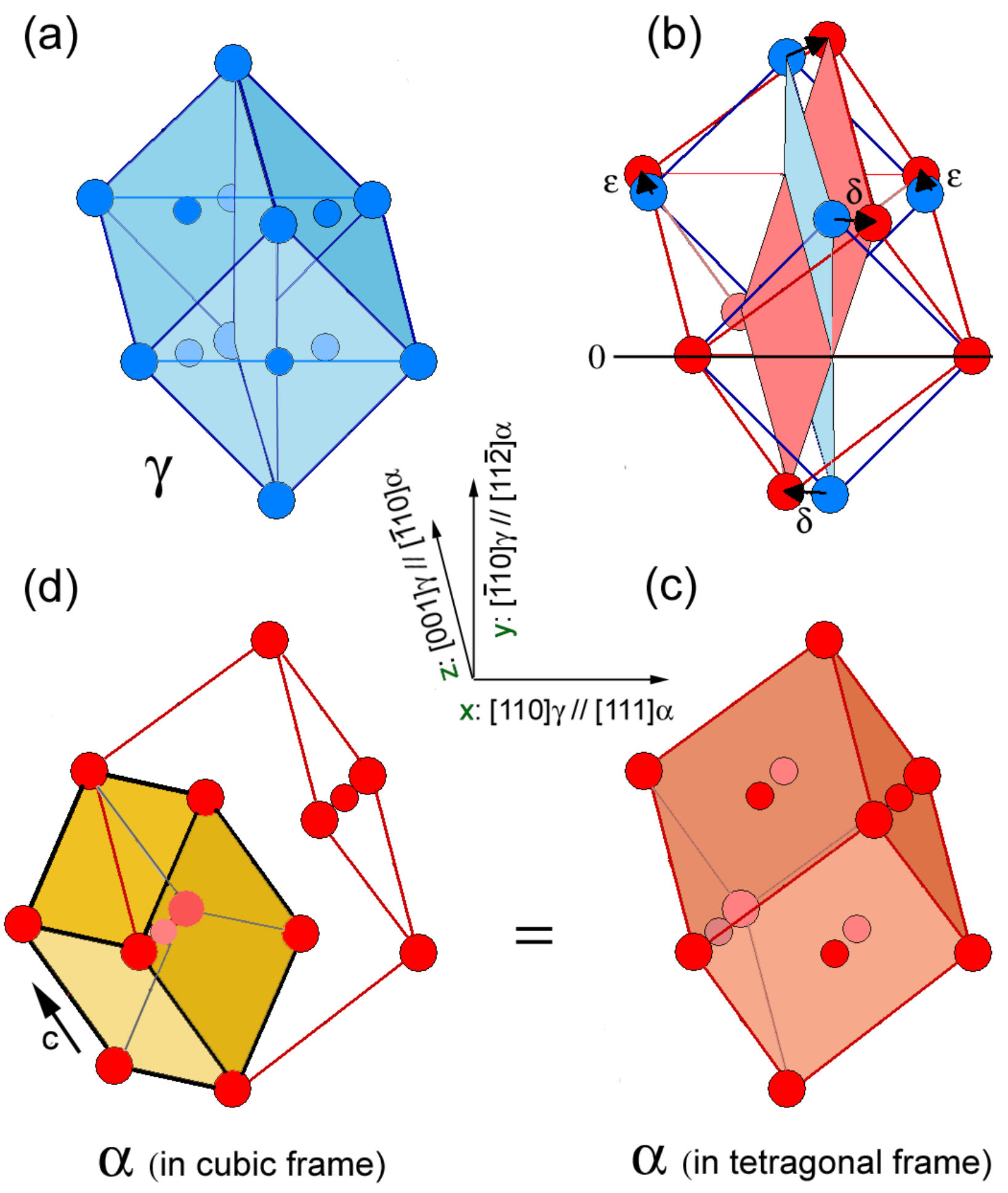

*Fig. 11. Fcc-bcc transformation by Pitsch distortion in 3D representation. (a) fcc cubic lattice lying with the $(110)_\gamma$ plane in vertical position, (b) Pitsch distortion with the x = $[110]_\gamma$ = $[111]_\alpha$ neutral line in horizontal position and marked by "0", (c) bcc crystal in a tetragonal frame after distortion, (d) same crystal in its basic cubic reference lattice (not well rendered y the perspective).*

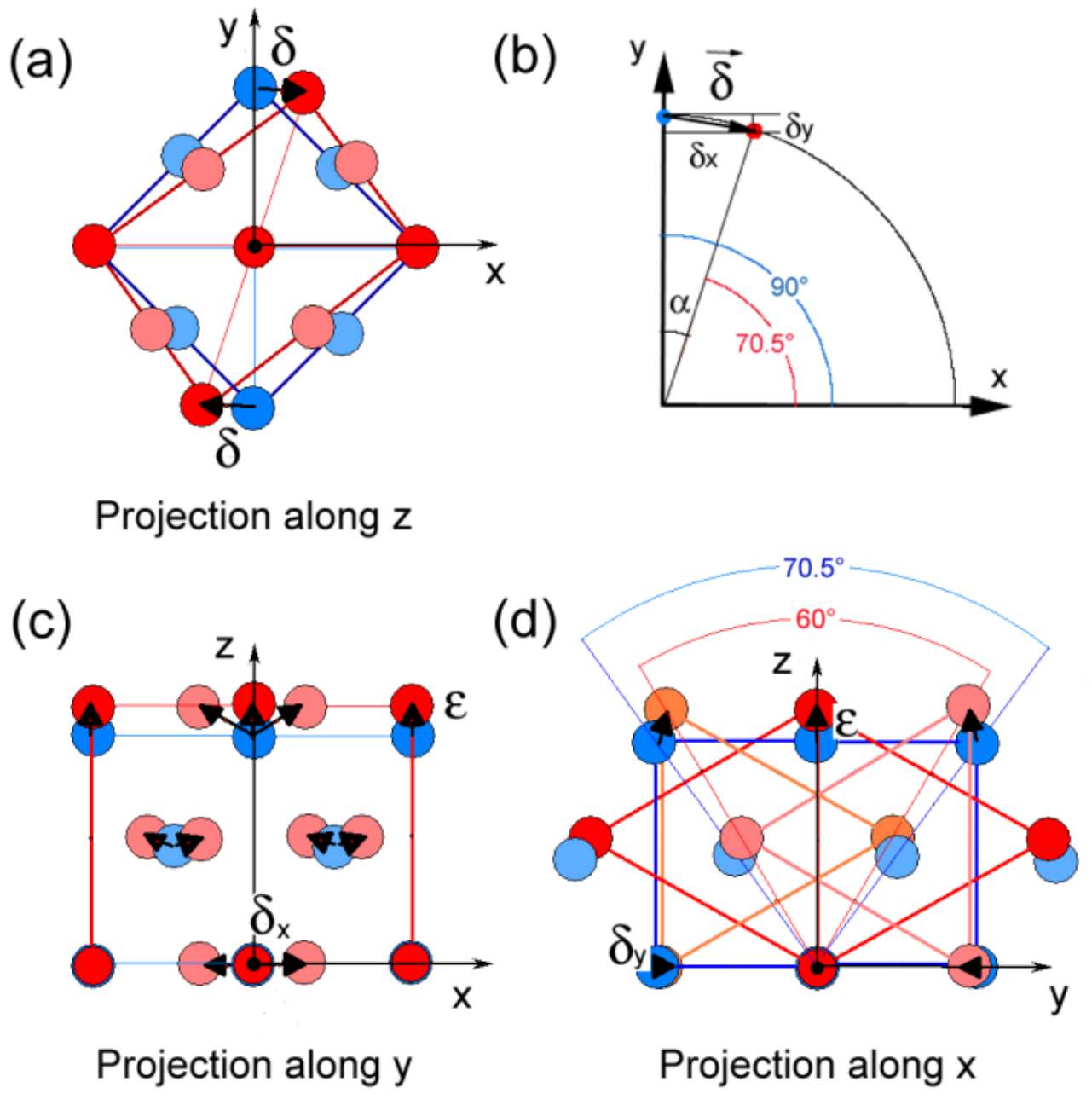

*Fig. 12. Fcc-bcc transformation by Pitsch distortion (see previous figure) viewed in projection along three axes: (a) and (b) z axis, (c) y axis, and (d) x axis.*

We consider now the distortion of the $(110)_\gamma$ plane into the $(111)_\alpha$ plane, viewed perpendicularly along the x axis (Fig. 12d). The angle of 70.5° between the $[1\bar{1}1]_\gamma$ and $[\bar{1}11]_\gamma$ diagonals of the $(110)_\gamma$ rectangle is distorted into the angle of 60° between the $[01\bar{1}]_\alpha$ and $[10\bar{1}]_\alpha$ diagonals of the $(111)_\alpha$ triangle. In addition the $[001]_\gamma$ direction is unrotated but is transformed into $[\bar{1}10]_\alpha$ direction by a dilatation of ratio $\varepsilon = 2/\sqrt{3}$ calculated from equation (3).

The Pitsch distortion matrix on the $B_1 = (x,y,z)$ basis is therefore:

$$D_{/B_1} = \begin{bmatrix} 1 & \delta_x & 0 \\ 0 & 1+\delta_y & 0 \\ 0 & 0 & \varepsilon \end{bmatrix} = \begin{bmatrix} 1 & 1/3 & 0 \\ 0 & \sqrt{8}/3 & 0 \\ 0 & 0 & 2/\sqrt{3} \end{bmatrix} \tag{4}$$

In the reference coordinate basis $B_0$ (of vectors $[100]_\gamma$, $[010]_\gamma$, $[001]_\gamma$) the base $B_1$ is given by the transformation matrix

$$[B_0 \to B_1] = \begin{bmatrix} 1/2 & -1/2 & 0 \\ 1/2 & 1/2 & 0 \\ 0 & 0 & 1 \end{bmatrix}, \text{ for which the inverse is } [B_1 \to B_0] = \begin{bmatrix} 1 & -1 & 0 \\ 1 & 1 & 0 \\ 0 & 0 & 1 \end{bmatrix}$$

The Pitsch distortion matrix in the reference coordinate base $B_0$ is therefore:

$$D_{/B_0} = [B_0 \to B_1] D_{/B_1} [B_1 \to B_0] = \begin{bmatrix} 1-\delta_b & \delta_b & 0 \\ -\delta_a & 1+\delta_a & 0 \\ 0 & 0 & \varepsilon \end{bmatrix} \quad (5)$$

with $\delta_a = (\delta_{x+}\delta_y)/2$ and $\delta_b = (\delta_{x-}\delta_y)/2$. $D/_{B0}$ is given numerically by:

$$D_{/B0} = \begin{bmatrix} \frac{2+\sqrt{8}}{6} & \frac{4-\sqrt{8}}{6} & 0 \\ \frac{2-\sqrt{8}}{6} & \frac{4+\sqrt{8}}{6} & 0 \\ 0 & 0 & 2/\sqrt{3} \end{bmatrix} \quad (6)$$

One can check that the determinant of the matrix D (whatever the chosen basis) is $\det(D) = 2\left(\frac{3}{2}\right)^{-3/2}$, as expected for a transformation between bcc and fcc phases in relationship by a hard sphere model, because the theoretical ratio of the atomic volumes between the phases is $r = \frac{a_\alpha^3/2}{a_\gamma^3/4} = 2\left(\frac{3}{2}\right)^{-3/2}$ according to equation (3).

The eigenvalues of D are real numbers $d_1 = 1$, $d_2 = \sqrt{8}/3 \approx 0.943$, and $d_3 = \varepsilon = 2/\sqrt{3} \approx 1.155$, with corresponding normalized eigenvectors $[1/\sqrt{2}, 1/\sqrt{2}, 0]_\gamma$, $[\sqrt{2/3}, \sqrt{1/3}, 0]_\gamma$, and $[001]_\gamma$ respectively. This means that the matrix D can be diagonalized by writing it in the basis $B_d$ formed by the eigenvectors:

$$D_{/B_d} = \begin{bmatrix} \frac{1}{\sqrt{2}} & \sqrt{\frac{2}{3}} & 0 \\ \frac{1}{\sqrt{2}} & \sqrt{\frac{1}{3}} & 0 \\ 0 & 0 & 1 \end{bmatrix}^{-1} D_{/B0} \begin{bmatrix} \frac{1}{\sqrt{2}} & \sqrt{\frac{2}{3}} & 0 \\ \frac{1}{\sqrt{2}} & \sqrt{\frac{1}{3}} & 0 \\ 0 & 0 & 1 \end{bmatrix} = \begin{bmatrix} 1 & 0 & 0 \\ 0 & \frac{\sqrt{8}}{3} & 0 \\ 0 & 0 & \frac{2}{\sqrt{3}} \end{bmatrix} \quad (7)$$

The differences between the diagonal values and unity give the principal strain values of 0%, -5.8 % and +15.5%. It is important to notice that the average of the absolute values is far lower than with the -20%, +12% and +12% values obtained with the Bain distortion, or even lower than with the two step mechanism [34]. Indeed, the average strain is only 7.1% with Pitsch, whereas it is 10.6% with the two-step model and 14.6% with Bain. Moreover, it is worth noting that, as Bain, no shuffle (movements of atoms inside the lattice) is required in the Pitsch distortion.

### 3.2 Pitsch distortion and the closing rotation A

From this analysis, Pitsch distortion seems to be the best candidate ever found to explain the fcc-bcc transformation. However, Pitsch OR is only a point on the continuous A and B paths, the average orientation being KS or NW. May Pitsch be the starting point of the continuous paths and may it explain them without other ad-hoc parameters or mechanisms? Even if difficult to prove it unambiguously, we will try to show in the following sections by simple physical arguments that the answer is very probably yes.

Let us consider the Pitsch distortion in projection along the neutral line x = $[110]_\gamma = [111]_\alpha$ (Fig. 13d). We represent $(110)_\gamma$ plane by an isosceles triangle (and not by a rectangle as in the previous section) and the $(111)_\alpha$ plane by an equilateral triangle. The angle of 70.5° is distorted into the angle of 60° of the $(111)_\alpha$ triangle. It is reasonable to imagine that the accommodation of such a distortion is completely accommodated by the γ matrix in which the α nucleus in Pitsch OR forms. This implies that on each face of the triangles, the fcc matrix lattice is progressively rotated from 0° far from the γ/α interface to 5.25° at the γ/α interface. This rotation is exactly the closing rotation A. What could be the physical mechanism of such rotations? Disclinations are a good description at mesoscopic scale. Disclinations were first introduced by Volterra in 1907 [42] who considered at that time two types of dislocations: rotational dislocations (disclinations) and translational dislocation (simply referred as dislocations nowadays). The strength of a disclination is given by an axial vector ***w*** (called Frank vector) encoding the rotation needed to close the system, such as the strength of a dislocation is given by its Burgers vector **b** encoding the translation needed to close the system (Burgers circuit). If dislocations constitute a fundamental part of metallurgy, disclinations remain confidential with few groups in world interested in them, mainly Romanov in Russia [43], Friedel and Kleman in France [44], and applications limited up to know to highly deformed metals by cold-work [45] or mechanical milling [46]. The 60° to 70.5° distortion associated to the neutral line x can be appropriately described by a wedge disclination of Frank axial vector $\boldsymbol{w_{2A}}$ = (-10.5°, $[110]_\gamma$), and can be decomposed into two closing wedge disclinations $\boldsymbol{w_A}$ = (-5.25°, $[110]_\gamma$) on each face of the α nucleus. A microscopic scale, the wedge disclinations can result from pile-ups on the $(\bar{1}10)_\gamma$ plane of edge sessile dislocations of line $[110]_\gamma$ and Burgers vectors **b** = $[\bar{1}10]_\gamma$ lying on the $(001)_\gamma$ plane, as symbolically represented in Fig. 14 (adapted from [43]). Such configurations of dislocations can be created by the accommodation of the martensite transformation or by prior plastic deformation of austenite. In that case, they could be associated to Lomer-Cottrel locks [47][48]. By simplification, the dissociation into Shockley partials introduced by Cottrel is not taken into account. The glide planes $(\bar{1}11)_\gamma$ and $(1\bar{1}1)_\gamma$ intersect into the x = $[110]_\gamma$ line. The Lomer locks can be obtained by two ways: (a) the ½ $[0\bar{1}1]_\gamma$ dislocation lying on the former plane combine with the ½ $[10\bar{1}]_\gamma$ dislocation lying on the latter plane to form a ½ $[1\bar{1}0]_\gamma$ dislocation, or (b) the ½ $[101]_\gamma$ dislocation lying on the former plane combine with the ½ $[0\bar{1}\bar{1}]_\gamma$ dislocation lying on the latter plane to form also a ½ $[1\bar{1}0]_\gamma$ dislocation. Both cases lead to a release of energy, and to edge dislocations on the $(001)_\gamma$ plane, which is not a slip plane for γ lattice, as illustrated in Fig. 14d.

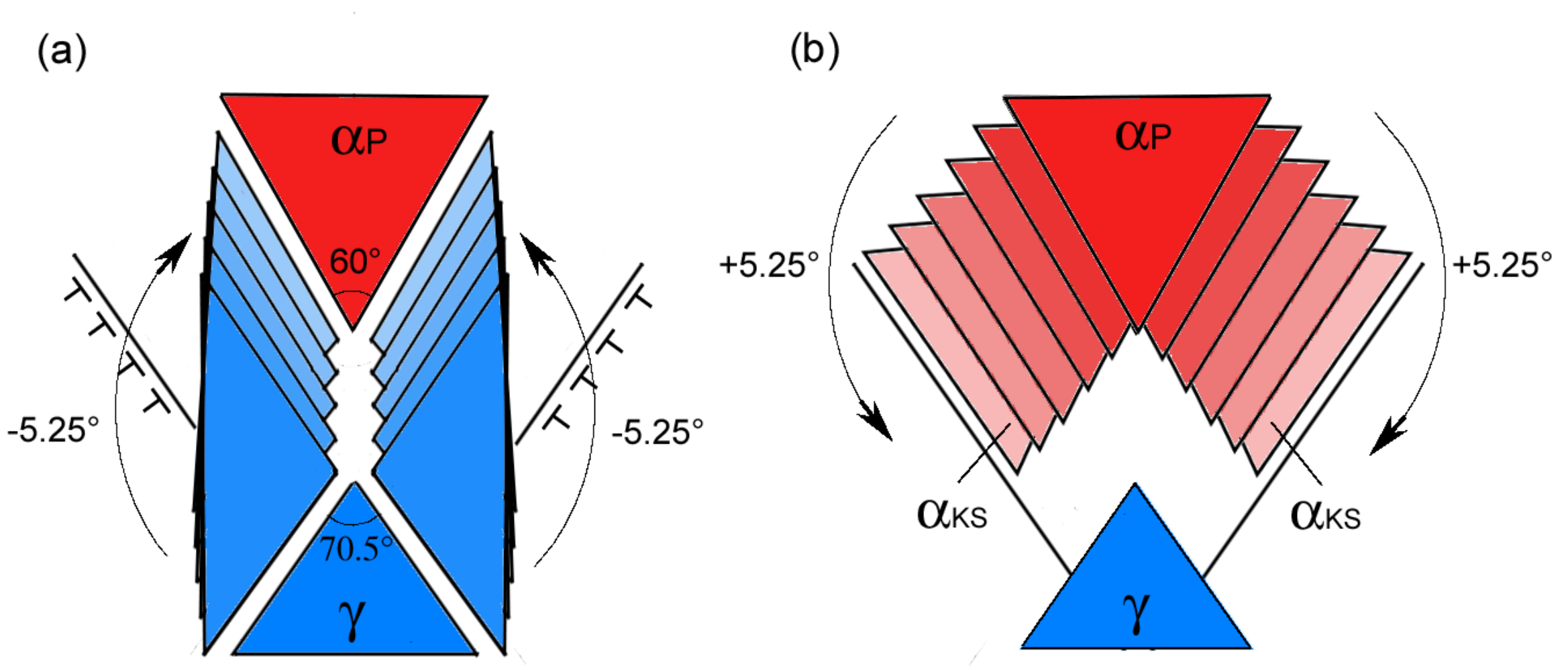

*Fig. 13. Explanation of the continuous rotation A between the Pitsch OR to the KS OR. (a) Nucleation of a $\alpha_P$ variant (in red) by Pitsch distortion and deformation of the γ surrounding matrix(in blue) induced by the γ→α transformation and accommodated by pile-up of dislocations creating the progressive rotation of 10.5°/2 on each side of the $\alpha_P$ nucleus. (b) Growth of the $\alpha_P$ variant into the deformed γ matrix by Pistch distortion (and local Pitsch OR with the matrix). In the reference frame of austenite far from the transformation, the nucleus is in Pitsch OR and its grown parts in KS OR forming two KS variants linked by the 2A operator.*

After the stage of α nucleation by Pitsch distortion, we can imagine that the transformation continues and martensite grows by the same Pitsch mechanism. However, now, the surrounding γ matrix is deformed and rotated, such that even if locally Pitsch OR is respected, the α martensite crystals appear to be progressively rotated globally in respect to the austenite matrix far away. The rotations are those created in the γ matrix by the transformation itself during the nucleation, they are the closing rotations A on both faces of the α nucleus. At the end of the process, as shown in Fig. 13b, even if the Pitsch distortion is the true mechanism, the initial nucleus variant in Pitsch OR is transformed into 2 variants derived from the Pitsch variant by a rotation A of +5.25°, i.e. into 2 variants in KS OR in reference to austenite, as illustrated in Fig. 9. Of course, one could ask why the formation of martensite by Pitsch distortion during growth does not create another deformation field of the γ surrounding matrix and an end-less process and infinite rotation. Even if that question is not yet solved, we believe that the KS OR constitutes a perfect locking configuration due to the parallelism of both close packed planes and directions.

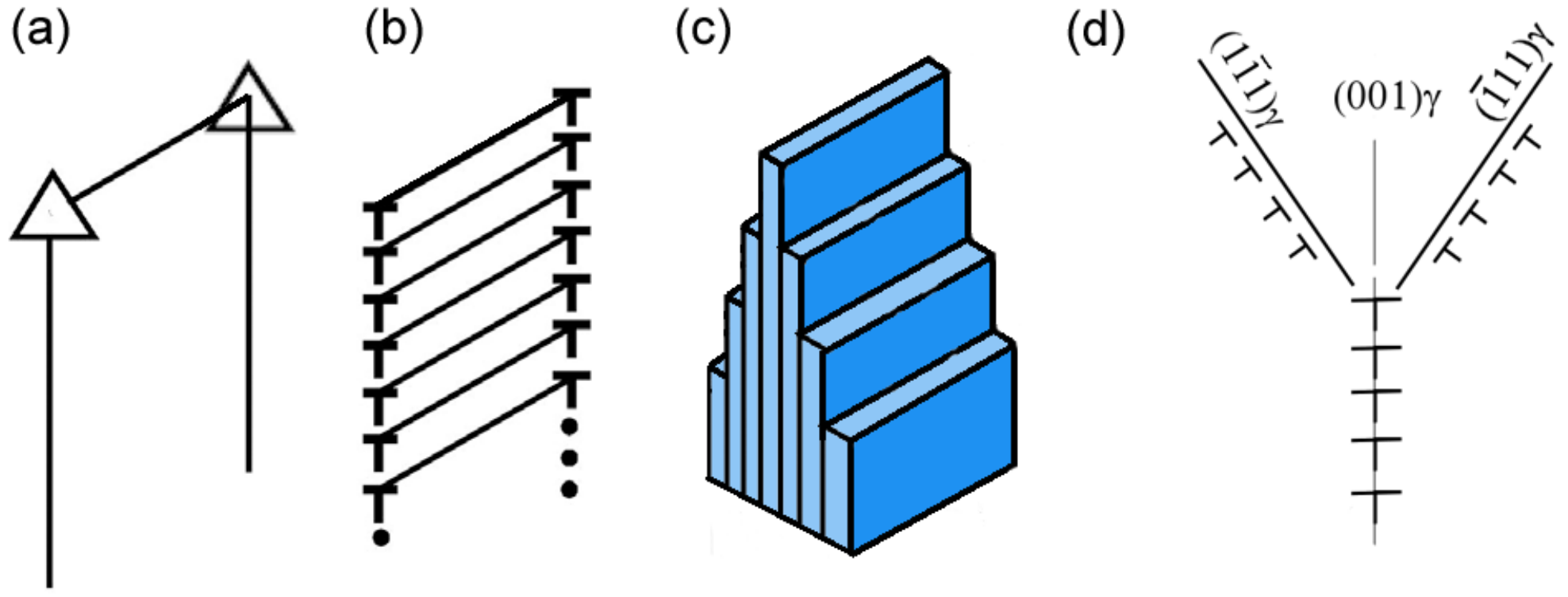

*Fig. 14. Wedge disclination viewed (a) by its symbolic representation, (b) as a pile-up of edge dislocations, (c) as series of additional planes, freely adapted from [43]. (d) Creation of the dislocation pile-up by Lomer-Cottrel lock.*

### 3.3 Pitsch distortion and the closing rotation B

Is it possible to explain the closing rotation B with similar arguments? At first sight the answer seems no, because none of $\{111\}_\gamma$ planes is parallel to a $\{110\}_\alpha$ plane with Pitsch OR. Only the low index plane $(001)_\gamma$ plane is transformed into the $(\bar{1}10)_\alpha$ plane (Fig. 15a). If one consider the $\{111\}_\gamma$ planes and the parallelism condition, it seems better start directly from KS to close two variants into one NW by rotation B (Fig. 8). However, we kept in mind the Boger and Burgers model that showed possible to transform by a simple shear on a $(111)_\gamma$ plane a $(\bar{1}11)_\gamma$ plane into a $(110)_\alpha$ plane. Could something similar be possible with the Pitsch distortion? Since we found difficult to figure out the effect of the Pitsch distortion on the $\{111\}_\gamma$ planes, we decided to calculate it with the matrix given in equation (6). For the four $\{111\}_\gamma$ planes, we determined the image of their normal and the images of the three $\{110\}_\gamma$ edges of the triangle. For three $\{111\}_\gamma$ planes, nothing special happened, but for the $(\bar{1}11)_\gamma$ plane the result is fascinating: of course the neutral x = $[110]_\gamma$ edge is unchanged, the $[101]_\gamma$ edge becomes ~[0.80,-0.14, 1.15]$_\gamma$ which makes an angle of 70.5° with x, and the $[01\bar{1}]_\gamma$ becomes ~[0.19, 1.14,-1.15], which makes an angle of -54.7° with x. The $[\bar{1}11]_\gamma$ normal becomes [-1.15, 1.15, 0.94]$_\gamma$ which makes an angle of 5.25° by a rotation around the [110]$_\gamma$ axis, i.e. by the rotation A. This proves that the effect of the Pitsch distortion on the $(\bar{1}11)_\gamma$ plane is (a) a rotation A, which gives the intermediate KS parallelism condition, and (b) a distortion that transforms it into a $(110)_\alpha$ plane, as shown in Fig. 15b and c, which is exactly the result expected! It is then possible to imagine for the distortion of the $(\bar{1}11)_\gamma$ plane a scenario similar to the one described previously for the $(110)_\gamma$ plane. The closing rotation is now rotation B and the pair of associated wedge disclinations are $\boldsymbol{w_B}$ = (-5.25°, $[\bar{1}11]_\gamma$). Such disclination can probably be obtained by a superposition of pairs or triplets of screw dislocations on the $(\bar{1}11)_\gamma$ plane.

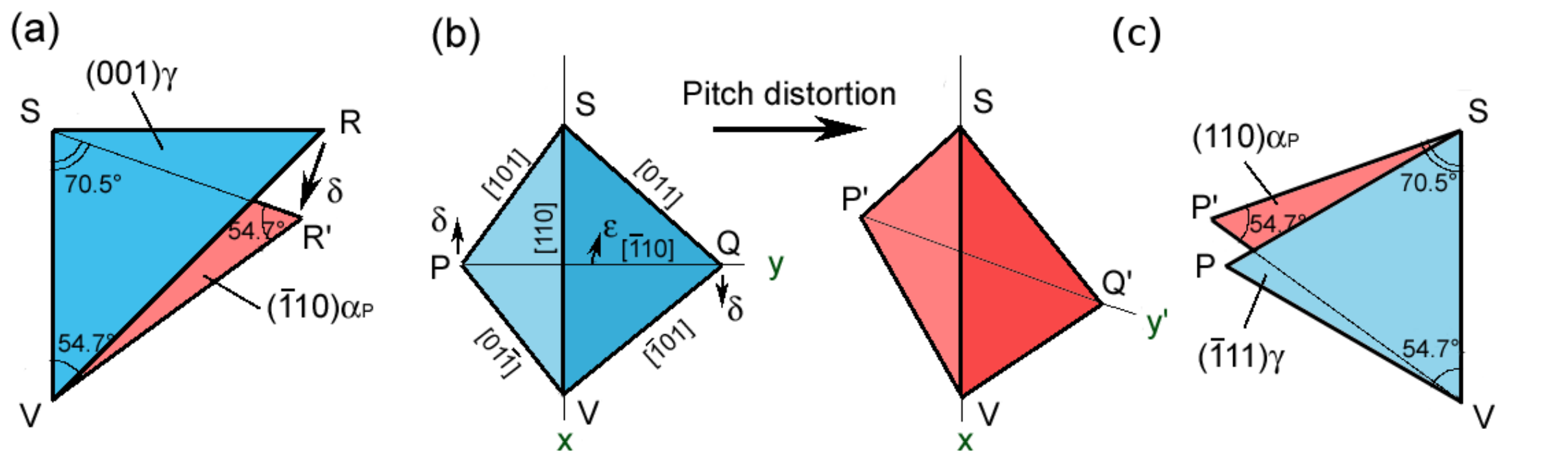

*Fig. 15. Effect of the Pitsch distortion on different planes: (a) on the $(001)_\gamma$ plane, (b) on the four $\{111\}_\gamma$ planes viewed in 3D, and (c) on the $(\bar{1}11)_\gamma$ plane. The $\gamma$ planes are in blue and the $\alpha$ planes in red.*

## 3.4 The Pitsch distortion and the A and B continuous paths

This global analysis shows that the Pitsch OR is a very good candidate to crack the nut formed by the close NW-KS-Pitsch structure of variants. The fascinating continuity between the ORs and the crystallographic intricacy between the CP directions and planes appear as a natural consequence of only one mechanism, the Pitsch distortion itself. For the first time, a simple physical scenario of atomic displacements of the martensitic transformation is proposed. Let us resume:

The α martensite nucleus appears by Pitsch distortion in Pitsch OR. This nucleation can be favoured by prior plastic deformation of austenite as it will be discussed in the next section.

The formation of this nucleus creates incompatibilities with the surrounding $\gamma$ austenitic matrix. These incompatibilities are the rotational misfits 2A and 2B, and they are accommodated in the $\gamma$ matrix by disclinations, probably constituted of pile-ups of edge and screw dislocations respectively. In other words, the $\gamma$ matrix has been continuously rotated by the two rotations A and B with angles varying from 0° far from the nucleus to 5.25° close the $\gamma/\alpha$ interface. During the martensite growth, the $\gamma \rightarrow \alpha$ transformation obeys the same Pitsch distortion, but now inside a rotated $\gamma$ matrix. Far far from the Pitsch nucleus, the new $\alpha$ variants are misoriented of 5.25° by rotations A and B, and thus appear in KS and NW ORs. These two ORs seem to act as locking configurations. In this scenario, the martensite transformation modifies the orientations of the variants by its own deformation field.

It is now possible to complete the CPP-CPD diagram introduced by Miyamoto *et al.* [12] to represent the deviation between the OR determined experimentally and the reference KS OR, with two angles $\theta_{CPP}$, the angular deviation between the CPPs $(111)_\gamma$ and $(110)_\alpha$, and $\theta_{CPD}$, the angular deviation between the CPDs $[110]_\gamma$ and $[111]_\alpha$. The Pitsch OR and the transformation path with its arrow can be added in the figure: the martensitic variants starts from a Pitsch OR, and are continuously reoriented by the deformation field created by the Pitsch distortion toward the KS and NW ORs, as shown in Fig. 16. The path A followed by B noted (A+B), which can be obtained by using KS and applying A($a$) with $a \in [0,5.25°]$ and B($b$) with $b \in [0,5.25°]$, supposes that the mechanisms are dissociated, which is not the case because both A and B result from the same mechanism. More realistic paths are combinations of the two, such as AxB = A($c$) + B(5.25°-$c$) with $c \in [0,5.25°]$ which gives the straight line from Pitsch OR to NW OR. Polynomial compositions, which give curves between Pitsch OR and NW OR, are also possible. The exact shape should depend on the mechanical properties of the austenite. The simulations presented in Fig. 10 were obtained with the A+B and AxB paths; they show that all the combinations of A and B occur during the variant reorientation in the austenite strain field. It can be noticed that the average Myamoto OR [12] is close to the barycentre of the P-KS-NW triangle, such as one could expect from an average OR obtained on a bundle of curved paths covering the surface of the triangle. The structure of the KS variants with their packets corresponding to Fig. 6 can now be represented with their starting nucleus Pitsch variants, their ending NW and KS variants, and the continuous paths (represented by A+B paths for sake of simplicity), as shown in Fig. 17.

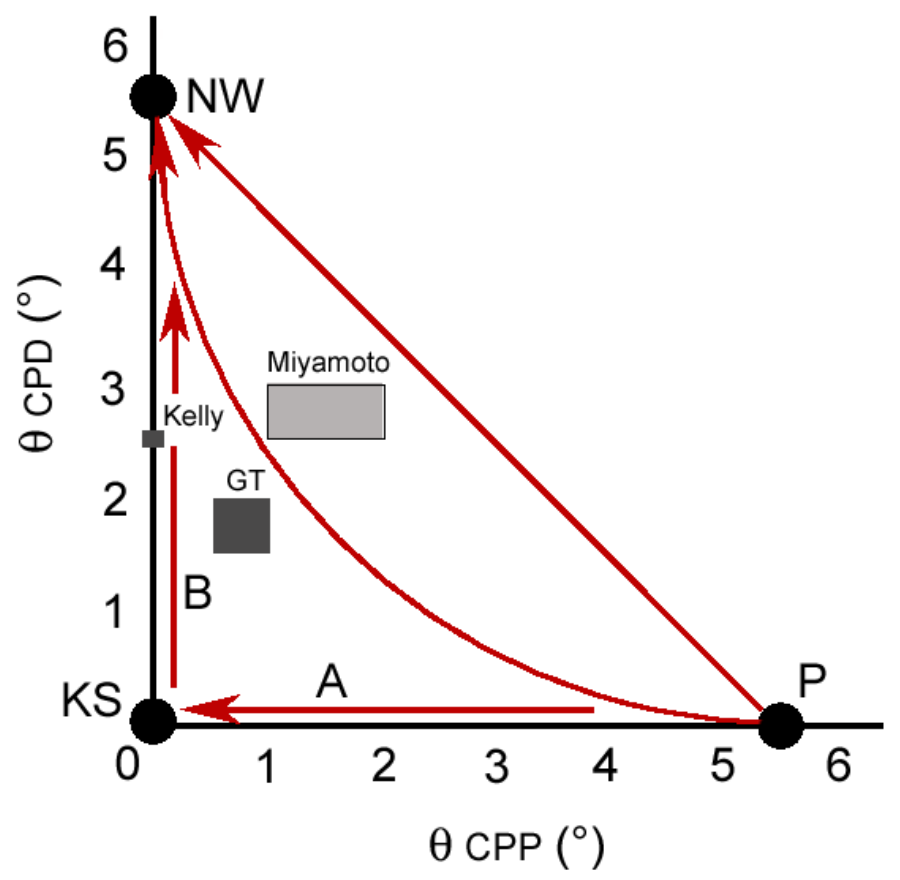

*Fig. 16. CPP-CPD diagram of martensitic transformation. The OR of the martensitic transformation is represented by its deviation from the KS OR. The angular deviation between the CPPs $(111)_\gamma$ and $(110)_\alpha$ is noted $\theta_{CPP}$ (x axis) and the angular deviation between the CPDs $[110]_\gamma$ and $[111]_\alpha$ is noted $\theta_{CPD}$. The one-step model explains the continuous path from the Pitsch OR to the KS and NW OR. The Kelly, GT and Myamoto ORs are also indicated.*

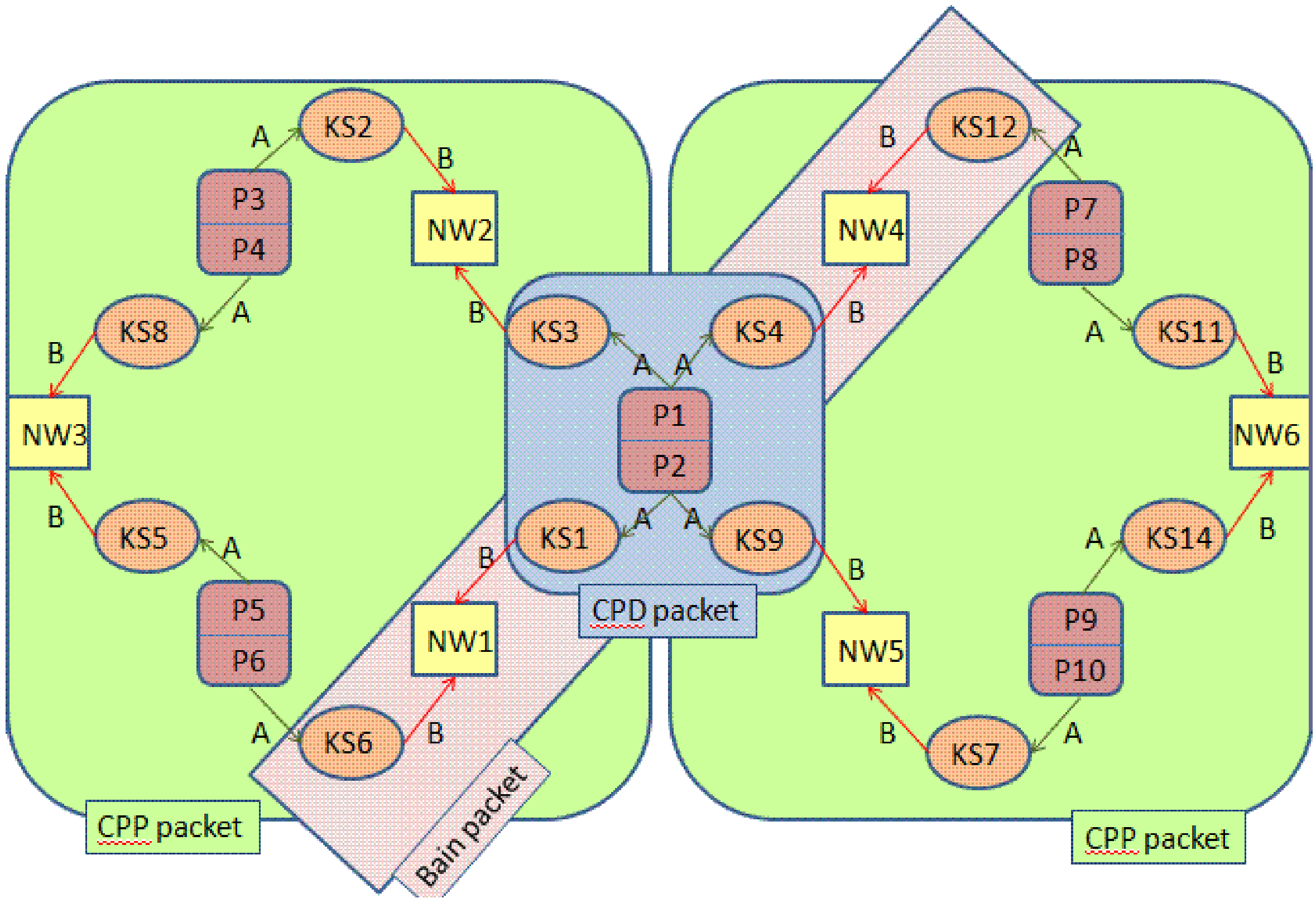

*Fig. 17. Schematic representation of the continuous paths between the twinned Pitsch nuclei (noted P1/P2, P3/P4 etc), the intermediate KS variants obtained by rotation A, and the final NW variants obtained by rotation B. Actually, A and B act simultaneously because they result from the same Pitsch distortion as represented in the previous figure.*

# 4 Discussions by revisiting literature

## 4.1 Carbon content and tetragonality

It can be noticed that the interstitial octahedral sites of austenite, i.e. the 12 centers of the <100>$_{\gamma}$ edges and the center of the lattice, which are partially occupied by carbon atoms, are transformed by Pitsch distortion into octahedral sites of the bcc martensite, i.e. the 12 centers of the <100>$_{\alpha}$ edges and the 6 centers of the {100}$_{\alpha}$ faces. Therefore, the Pitsch distortion could explain, as well as Bain distortion, the fact that carbon atoms occupy the octahedral sites in the $\alpha$ Fe structure, despite smaller space than in tetrahedral sites. Moreover, as Bain, the occupation of octahedral sites may explain the tetragonal distortion of the martensite (bcc$\rightarrow$bct) along one of the <100>$_{\alpha}$ axis, which becomes the c axis of the bct structure at high carbon content. With Bain distortion, it is assumed that the c axis is the axis of compression, even if the link between the ordering of carbon atoms in some types of octahedral site has never been fully clarified. Whatever the details, the same explanation should hold for Pitsch distortion. Indeed, with both Bain and Pitsch the final <100>$_{\alpha}$ axes come from the distortion of the same directions of the parent $\gamma$ phase: two <100>$_{\alpha}$ axes come from the two perpendicular $[0\bar{1}1]_{\gamma}$ and $[01\bar{1}]_{\gamma}$ of the $(100)_{\gamma}$ plane, and the third <100>$_{\alpha}$ axis come from the $[100]_{\gamma}$ axis, which is therefore the natural candidate to be the c-axis, as presented in Fig. 11d.

## 4.2 Habit planes

The HPs were firstly determined in the 1930's from optical microscopy on martensite plates formed in monocrystalline austenite of known orientation (measured by X-ray diffraction), and they were given naturally in the reference frame of austenite. Later in the 1950's the HPs could be determined more precisely by TEM, in the austenite or martensite reference frame, but the custom made the searchers continue to report them in austenite. The HPs have been found to have exotic indexes such as $(225)_{\gamma}$, $(3,5,10)_{\gamma}$ etc. The PTMT was born to explain these strange HPs which do not correspond to usual $\gamma$ glide planes nor to the possible shear that could be deduced from the orientation relationship between the austenite and martensite (see section 1.1).

From a crystallographic point of view, trying to understand or predict HPs without understanding the mechanisms, as done in the PTMT, seems very tricky and limited. Since a mechanism based on Pitsch distortion is now proposed, is it possible to understand with simple arguments the HPs of martensite? We think it is. For martensite in low carbon steels with needle shapes along <111>$_{\alpha}$ [49] (thus without HP), the explanation is simple: they are elongated along the neutral line $[111]_{\alpha}$ = $[110]_{\gamma}$ which is the lowest strained direction of the transformation. For the exotic HP, we have drawn them in a stereographic pole figure and tried to find a correspondence with low index planes of martensite. We intended different solutions and found that the $\{112\}_{\alpha}$ planes with Pitsch OR give very satisfying result for the $\{225\}_{\gamma}$ HPs and are also quite close to the $\{259\}_{\gamma}$ and $\{3,10,15\}_{\gamma}$ HPs, as shown in Fig. 18a. The $\{135\}_{\gamma}$ HPs could appear as a $\{112\}_{\alpha}$ planes of martensite, but with KS OR, as shown in Fig. 18b. Only the $\{557\}_{\gamma}$ HPs remains not completely explained. Since the angle between the $(557)_{\gamma}$ and $(111)_{\gamma}$ plane is 9.4°, it is possible to imagine a link between them by the rotation 2A, but a coherent explanation is not yet found.

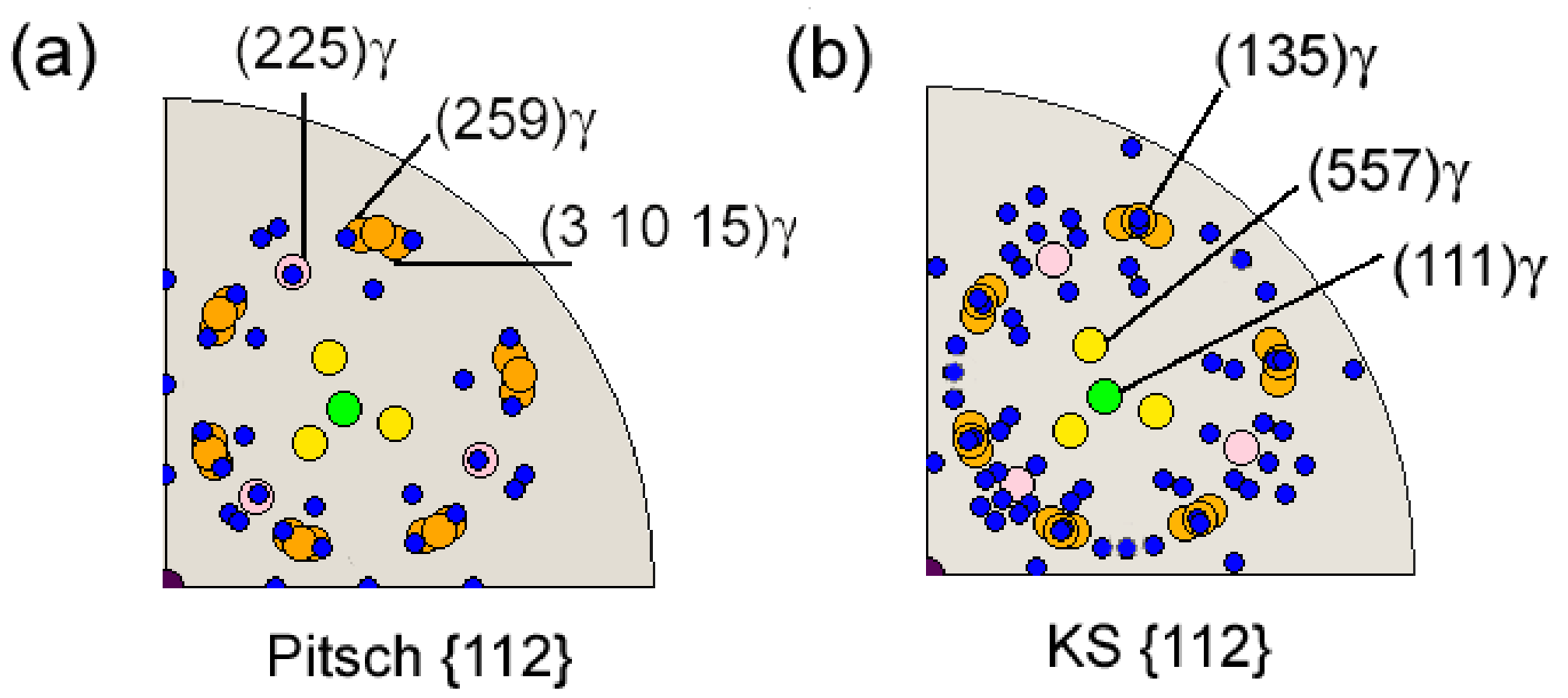

*Fig. 18. Stereographic projection of the most reported habit plane of martensite (large circles annotated in the reference lattice of the austenite), with (a) the $\{112\}_{\alpha}$ planes of the 12 Pitsch variants and (b) the $\{112\}_{\alpha}$ planes of the 24 KS variants (small blue disks). There is a good correspondence, except for the $(557)_{\gamma}$, planes.*

The $\{225\}_{\gamma}$ // $\{112\}_{\alpha}$ HPs are in perfect agreement with the Pistch distortion because they contain the neutral line. We can thus even precise which $\{225\}_{\gamma}$ and $\{112\}_{\alpha}$ HPs in their symmetrically equivalent families are correct in reference to the $[110]_{\gamma}$ // $[111]_{\alpha}$ neutral line: they are the $(2\bar{2}5)_{\gamma}$ and $(\bar{2}25)_{\gamma}$ , and the $(11\bar{2})_{\alpha}$, $(1\bar{2}1)_{\alpha}$ and $(\bar{2}11)_{\alpha}$ planes, as it will be detailed in the next sections.

### 4.3 Nucleation of martensite at intersection of glide planes

It has been recognized for a while that plastic deformation of austenite before quenching makes the martensite start temperature increases, which means that prior plastic deformation favors martensite formation. This strain-induced nucleation of martensite was also clearly shown to involve shear systems of austenite such as staking faults or ε plates [26]. These observations were the starting point of the Bogers and Burgers model of martensite transformation as explained in section 1.1.

The fact that the martensite tends to form at the intersection of two shear bands of austenite can be explained with the one-step theory. The intersection line of the two shear or glide bands is indeed the place of very high strain concentrations that help to distort the 70.5° of the $(110)_\gamma$ plane into the 60° of the $(111)_\alpha$ plane as described in section 3.2. The shear intersections could act as Lomer-Cottrel locks and sources of wedge disclinations that would reduce the energy gap between the γ and α phases, and therefore trigger the martensite nucleation. We can illustrate such a situation by looking at the TEM images of Shimizu and Nishiyama [50], as the one reported in Fig. 19a.

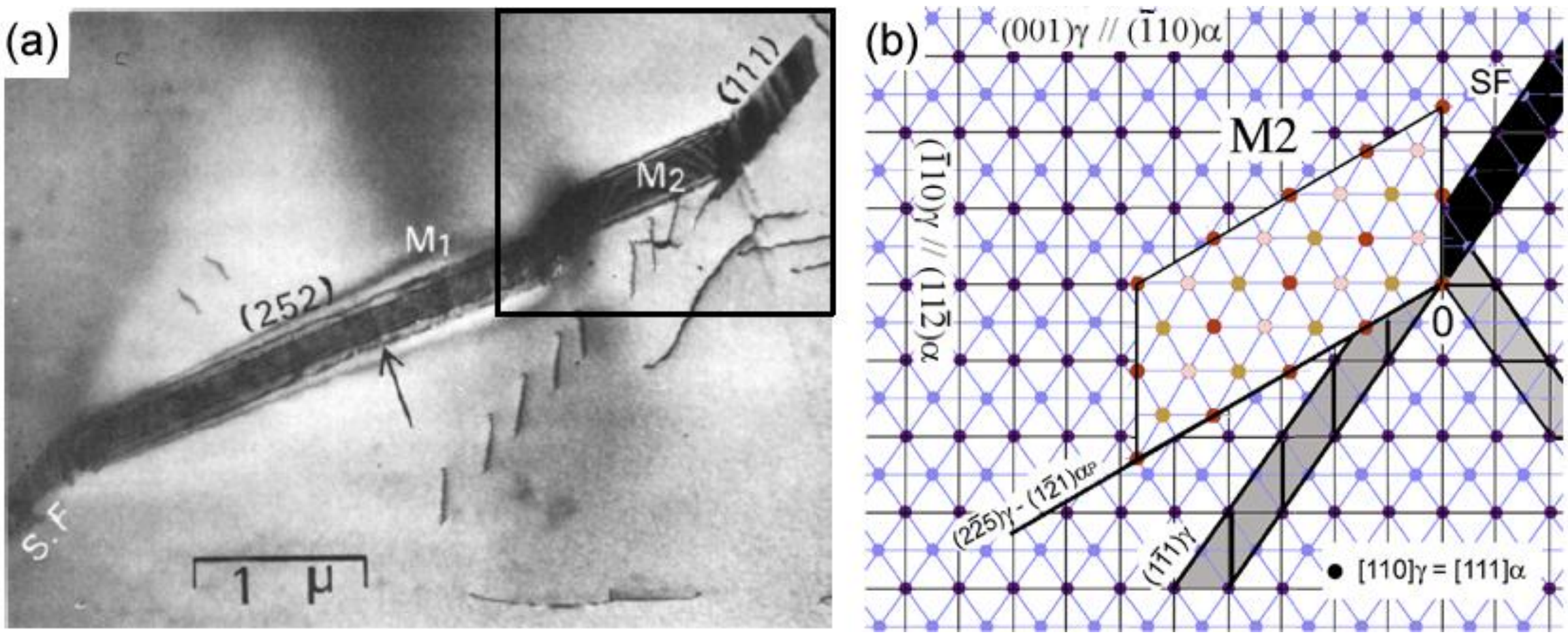

*Fig. 19. Formation of a martensite lath with $(2\bar{2}5)_\gamma$ habit plane. (a) TEM image from Shimizu and Nishiyama [50] showing at the up right corner the formation of a martensite $M_2$ at the junction of two $\{111\}_\gamma$ austenite glide planes. (b) Schematic presentation with the lattices of the phases and the indices used in the text. The $(1\bar{1}1)_\gamma$ glide plane is in grey with dislocations represented by the black vertical lines. This glide plane is transformed into a stacking fault (SF) indicated by the black band.*

This image is also discussed by Nishiyama in his book [18]. It is clear on the M2 part that the martensite nucleated at the intersection of two $\{111\}_\gamma$ glide plane (only faint trace of the right one is visible). The HP plane of the new formed martensite M2 is $\{225\}_\gamma$ and Nishiyama noticed that from all the equivalent $\{225\}_\gamma$ planes only those at 25° from the $\{111\}_\gamma$ glide plane can appear. He added "*this fact must be taken into account in the nucleation theories*". By keeping the reference frame used to describe the Pitsch distortion in section 3, that point can be understood as follows. Let assume that the TEM image was acquired with an electron beam close to the neutral line $[110]_\gamma$ // $[111]_\alpha$, the glide plane can be indexed as $(1\bar{1}1)_\gamma$ and the HP as $(2\bar{2}5)_\gamma$ , which should also be a $\{112\}_\alpha$ plane for the martensite in Pitsch OR (but no indication is given in ref. [50] or [18] on that point). The $(2\bar{2}5)_\gamma$ and $(1\bar{1}1)_\gamma$ planes make an angle of 25.24°. It can also be noticed that the $(1\bar{1}1)_\gamma$ glide plane is transformed into a stacking fault beyond the intersection point. Such an observation can be explained by considering the Lomer-Cottrel locks. For sake of simplicity we introduced them in section 3.2 without taking into account the dissociation of dislocations into Shockley partials. But actually, the $\frac{1}{2}[0\bar{1}1]_\gamma$ dislocation lying on $(\bar{1}11)_\gamma$ plane and the $\frac{1}{2}[10\bar{1}]_\gamma$ dislocation lying on the $(1\bar{1}1)_\gamma$ plane dissociate during combination, the former is split into 1/6 $[\bar{1}\bar{2}1]_\gamma$ and 1/6 $[1\bar{1}2]_\gamma$, and the latter into 1/6 $[21\bar{1}]_\gamma$ and 1/6 $[1\bar{1}\bar{2}]_\gamma$. The first members of each pair attract each other and combine according to

$$1/6\,[\bar{1}\bar{2}1]_\gamma + 1/6\,[21\bar{1}]_\gamma \rightarrow 1/6\,[1\bar{1}0]_\gamma \qquad (8)$$

The resulting edge dislocations of Burgers vector 1/6 $[1\bar{1}0]_\gamma$ and line x = $[110]_\gamma$ are sessile. They pile-up in the $(001)_\gamma$ plane and form the edge disclination. Even if not reported in literature to our knowledge, one can imagine that the residual Shockley partials 1/6 $[1\bar{1}2]_\gamma$ and 1/6 $[1\bar{1}\bar{2}]_\gamma$ continue to glide in their respective $(\bar{1}11)_\gamma$ and $(1\bar{1}1)_\gamma$ planes, transforming their glide plane into stacking faults, as observed in Fig. 19a (at least for the visible glide plane). A schematic drawing with lattice and atomic positions has been added in Fig. 19b, without taking into account the atom reorganization at the $\gamma/\alpha$ interface. Dissociations of perfect dislocations into Schokley partials associated to $(225)_\gamma$ martensite and stacking fault have already been observed by TEM in Fe-Cr-C alloys [51].

According to our approach, prior plastic deformation of austenite would increase the number of intersection of $\{111\}_\gamma$ glide planes which would act as nucleation sites for martensite for reasons explained previously. Moreover, as shown in Fig. 6a and in the scheme of Fig. 17, each site creates two Pitsch variants which are transformed into four KS variants, all sharing the common $[111]_\alpha$ // $[110]_\gamma$ neutral line, forming a CPD packet as detailed in section 0. We believe that the sets of variants produced in a strained Fe-Ni single crystal by martensite burst transformation and whose poles cluster about common $\langle 110 \rangle_\gamma$ directions [52] correspond to these CPD packets.

## 4.4 Midrib, twins and growth.

Some martensitic alloys sometimes have lenticular or butterfly shapes that exhibit a planar zone inside, often in their middle, called midrib. The formation of midribs has never been fully explained despite the numerous researches (a historical review can be found in the Nishiyama book [18] pp 43-47). It is agreed now that the midrib is the plane of initiation of the transformation, and the $\gamma/\alpha$ interface propagates laterally on either side in opposite directions [18]. Sometimes the boundaries of the midrib are sharp and the midrib can be considered as a thin plate [53][54]. Is has been noticed a gradual rotation between the midrib and the external $\gamma/\alpha$ interfaces [53][55], but such an observation remains unexplained. The midribs also often contain a high density of "twins" that were promptly viewed as mechanical twins assimilated to the "inhomogeneous lattice-invariant" deformation required by the PTMT (see section 1.1 and ref. [18][20][49] for examples). Even if this view is shared by many metallurgists, some features do not fit it. Let us for example consider the TEM image of Fig. 20a.

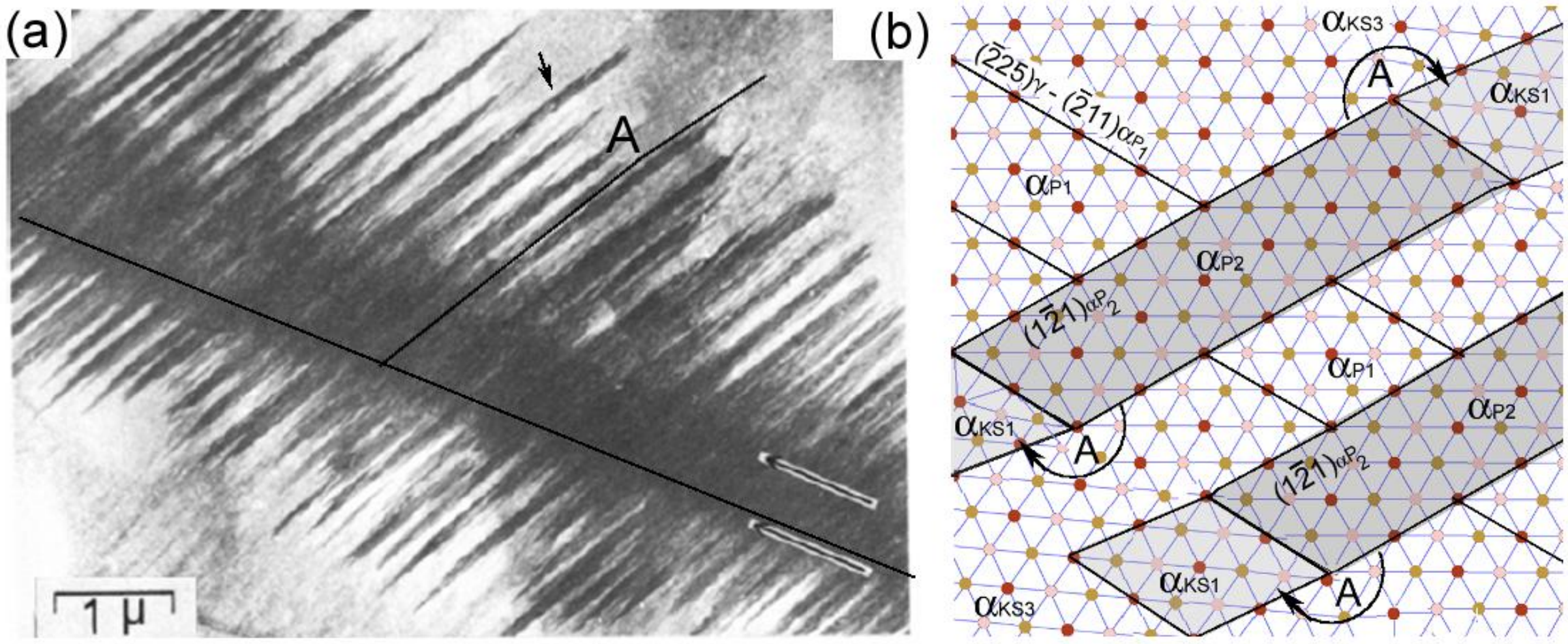

*Fig. 20. Midrib and « twins » inside a lenticular martensite lath. (a) TEM image from from Shimizu and Nishiyama [50]. The "twins" habit planes have an angle of 60° with the midrib and they are curved at the mark "A", also visible at the arrow tip. (b) Schematic presentation with the lattices of the phases. The "twins" are actually Pitsch variant $\alpha_2$ in twin orientation relationship with the main Pitsch variant $\alpha_1$ forming the midrib. The HP of both variants are $\{112\}_a$ planes and the curvature is in agreement with rotation A which results from the deformation of the surrounding matrix during the formation of the nucleus $\alpha_1$ at the midrib.*

Why twins stop inside the martensite and not cross it entirely? Moreover, an important detail is sometimes noticed but often ignored: the boundaries of the "twins" are not straight but generally slightly curved all in the same direction [50][51][56], which is unusual for mechanical twins. All these particular features can be explained within the one-step theory, if we consider that the midrib corresponds to the nucleus Pitsch variant ($\alpha_{P1}$) and that the "twins" do not result from a mechanical shear but from the martensite transformation itself. Indeed in each lenticular martensite only one "twin" orientation is observed among the four possible variants corresponding to the four rotations of 60° around the $<111>_\alpha$ directions, and this "twin" is actually the other Pitsch variant ($\alpha_{P2}$) sharing the same neutral line with the nucleus $\alpha_{P1}$. The $\alpha_{P1}$ - $\alpha_{P2}$ pair of Pitsch variants results from the Pitsch distortion: the $\delta$ part of tensor (4) can be obtained in two directions deduced by the $(11\bar{2})_\alpha$ mirror symmetry (Fig. 11b or Fig. 12a). In first rough approximation, by assuming that $\delta$ is a shear, the mirror is equivalent to inverse the direction of the shear vector. The two Pitsch variants $\alpha_{P1}$ and $\alpha_{P2}$ share the same x = $[111]_\alpha$ direction and are linked by a rotation of 60° around it. During nucleation of $\alpha_{P1}$, intricate nucleation of $\alpha_{P2}$ inside $\alpha_{P1}$ is probable because both crystals share a common lattice (the $\Sigma3$ coincidence site lattice). These pairs of twin related variants were observed by Pitsch in the martensite transformed thin TEM lamella [11]. As illustrated in Fig. 20b, after nucleation, the $\alpha_{P1}$ variant grows in a $(\bar{2}11)_\alpha$ habit plane, forming the midrib, and the $\alpha_{P2}$ in the $(1\bar{2}1)_\alpha$ habit plane, which is also the mirror plane of $\Sigma3$ misorientation. The two $(\bar{2}11)_\alpha$ and $(1\bar{2}1)_\alpha$ HPs have an angle of 60°. The rotation A, generated by the deformation field of the Pitsch distortion (section 3.2), acts differently on the $\alpha_{P1}$ and $\alpha_{P2}$ variants because of the difference of their orientations and habit planes. The $\alpha_{P1}$ midrib continues to grow laterally and its lattice is gradually rotated so that its orientation gets close to KS: $\alpha_{P1}$ has been transformed progressively into the $\alpha_{KS3}$ variant according to the scheme of Fig. 17. The martensite aquires a lenticular shape often asymmetric due to the rotation A. The $\alpha_{P2}$ "twins" continues to grow; the rotation A transforms them progressively into a $\alpha_{KS1}$ variant according to the scheme of Fig. 17, and also curves the $(1\bar{2}1)_\alpha$ boundary. This 5.25° curvature marked by a broken line in Fig. 20a, creates additional strains with the surrounding $\alpha_{P1}$ crystal, which quickly stops the growth process, as represented in Fig. 20b. The $\alpha_{P1}$, $\alpha_{P2}$, $\alpha_{KS3}$ and $\alpha_{KS1}$ variants belong to the same CPD packet.

## 4.5 Butterfly martensite

Among the wide variety of shapes and morphologies, the butterfly one is probably the most intriguing. Butterfly martensite is formed by two lenticular shape wings on two distinct $\{225\}_\gamma$ planes, and, as them, they can exhibit the same internal features, such as midrib, “twins”, $(110)_\alpha$ planar faults and serrations [57][58][59]. Sato *et al.* [60][61] could show recently by EBSD that butterfly martensite inside a wing is also gradually deformed by a rotation around the common $[111]_\alpha = [110]_\gamma$ axis and of maximum angle generally found between 5 and 10°, i.e. by the rotation A. There is another interesting point that can be noticed in the EBSD map reported in Fig. 21a: the two wings have the same orientation and the same internal gradient of orientations despite the fact they have different HPs. This result is in contradiction with PTMT which is based on one habit plane only. Let us analyse it with the “one-step” approach. The fact that the two wings of the butterfly martensite have similar size proves that they result from the same nucleation process. The model of nucleation proposed in section 4.3 seems to apply. Indeed, as shown in Fig. 21b, it is probable that the butterfly has first nucleated at the point noted “0”, which is the intersection of two glide planes. Only one Pitsch variant is nucleated. The $(002)_\gamma$ // $(\bar{1}10)_\alpha$ mirror symmetry of the austenite-Pitsch variant system makes the growth on the two $(\bar{2}25)_\gamma$ // $(\bar{2}11)_\alpha$ and $(2\bar{2}5)_\gamma$ // $(1\bar{2}1)_\alpha$ HPs equivalent and explains the two wings for a unique Pitsch variant. Moreover, on the butterfly martensite of Fig. 21a there is no midrib, which means that the growth occurs only to inner direction with the same rotation A for both wings whatever the HP. The growth becomes limited when the $(01\bar{1})_\alpha$ planes become parallel to $(1\bar{1}1)_\gamma$. Rotation A has broken the mirror symmetry and generates a KS variant. The case studied here corresponds to a monocristalline butterfly martensite in which the two $(\bar{2}25)_\gamma$ and $(2\bar{2}5)_\gamma$ HPs intersect into the neutral line $[111]_\alpha = [110]_\gamma$. The angle of the two wings is 58.99°. From literature it seems that there are many other configurations for the pairs of $\{225\}_\gamma$ HPs in which they do not intersect at a $<110>_\gamma$ line, but at $<120>_\gamma$ or other lines with angles ranging from 41° to 139° [58]. These cases are more difficult to understand; the pair of wings could correspond to a pair of Pitch variants with their own distinct neutral line, and their nucleation would not obey the mechanism described in section 4.3. Systematic precise EBSD studies on the pairs of wings could be helpful to solve this point.

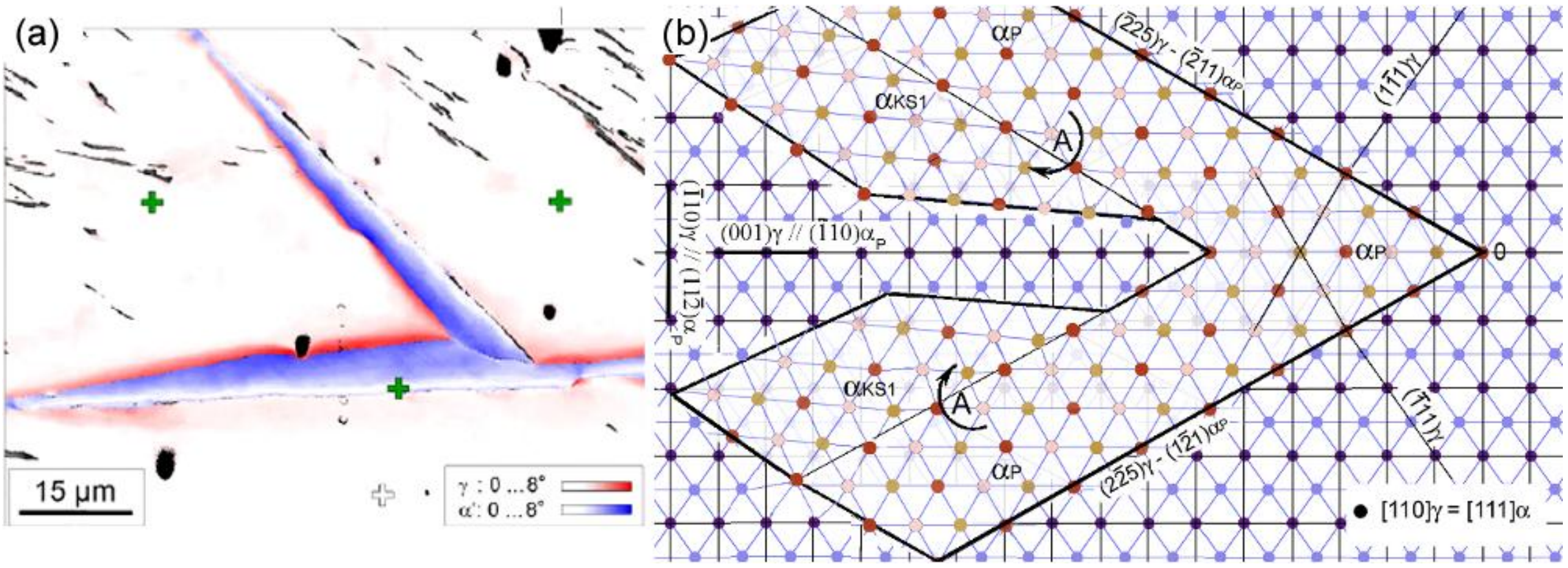

*Fig. 21. Butterfly martensite. (a) EBSD map showing the misorientations inside the martensite (in blue) and in the austenite matrix (in red), from Sato and Zaefferer [60].The colours are reversed in comparison to the ones chosen in our paper. (b) Schematic presentation with the lattices of the phases. Nucleation of a martensite variant in Pitsch OR at point noted "0", and growth, first along the $\{112\}_\alpha$ planes and latter by thickening of the wings and gradual rotation A .The angle between the plane $(2\bar{2}5)_\gamma$ and $(\bar{2}25)_\gamma$ is 58.99° and the angle between $(11\bar{2})_\alpha$ and $(2\bar{1}\bar{1})_\alpha$ is 60°. It is important to notice that both wings of the butterfly have the same initial Pitsch OR and the same internal continuous rotation A despite their different habit planes. The Pitsch OR is not reported in [60], but the authors studied only the OR at the $\gamma/\alpha$ interface; it is possible that Pitsch OR is inside the wings along the frontier between the white and blue parts.*

### 4.6 Internal and edge defects

In our approach, many features of martensite result from the Pitsch distortion and its effect on the surrounding austenite (rotations A and B). At nucleation stage, we have seen in section 4.4 that the "twins" in the midrib are actually pairs of Pitsch variants, a configuration probably necessary to reduce the strains induced by the transformation. At growth stage, following the same idea, the many $(110)_\alpha$ // $(111)_\gamma$ planar defects observed in lenticular martensite [50][51][56] and butterfly martensite [57][58][59] can probably be explained as follows: the deformation of austenite in the surrounding of a nucleated martensite is accommodated by the disclinations A constituted of dislocations on a specific $(111)_\gamma$ plane (section 3.2). When the austenite is in its turn transformed into martensite in local Pitsch OR with it, the $(111)_\gamma$ planar defects are probably transformed into $(110)_\alpha$ defects by "inheritance". Similarly the many sets of parallel screw dislocations with Burgers vector ½ $[111]_\alpha$ // ½ $[110]_\gamma$ [51] could result from the accommodating disclinations B (section 3.3) and an inheritance mechanism during the martensitic transformation.

Serrations are small notches sometimes visible on the edge of martensite [3][56]. They were observed to be linked to some $(111)_\gamma$ planar defects but their origin is not understood. "*The serrations can correspond to the termination of growth in local sections of the plates due to constraints in the matrix, or to the nucleation of side plates of other habit plane variants*" [56]. This seems contradictory. If one can easily understand that the $(111)_\gamma$ planar defects acts as barriers and block the martensite growth, the fact that they could help the nucleation is more difficult to comprehend with classical theories. With the one-step theory, since we have shown that the intersection of two glide planes can trigger the martensite Pitsch distortion (section 4.3), the striations can be imagined as regeneration of the martensite growth, the dislocations on the $(111)_\gamma$ planar defects feeding the disclinations necessary to the martensite transformation. These defects probably stop the growth of the martensite plates for a short moment and once the intersection points is sufficiently filled of dislocations to create the disclination, the martensite grow again on its main habit plane, or on another $(225)_\gamma$ HP as in butterfly martensite.

### 4.7 Perspectives

The one-step model based on Pitsch distortion is self-consistent and seems to be in agreement with many observations reported in literature. A lot of microstructural features can now be re-investigated in light of this theory in order to confirm, modify or refute it. Among the possible experimental studies, EBSD investigations of the orientations and gradients of rotations in the pairs of wings of butterfly martensite seem interesting, in order to go further than what was already done in one wing [60]. We would like to stress here that many cares should be given to draw the $\gamma/\alpha$ interphase boundaries in EBSD maps with NW and Piscth ORs because they are both represented by the same misorientation, i.e. the angle/axis rotation of 45.98° around <8.3 20.1 97.6> [30] and can therefore be mistaken. New techniques allowing the increase of the spatial resolution in the orientations mapping are also promising, such as automatic TEM diffraction indexing, called ASTAR [62] and the electron forward scatter (ESFD) technique, also called t-EBSD [63]. Precise measurement of strain accommodation in austenite matrix surrounding martensite has already been developed recently by Myamoto et *al.* [64], but their results, as those that could be obtained in further studies, could be re-investigated and discussed with the one-step model. 3D EBSD with special data treatments [65] can also be used to determine automatically the HPs in order to better understand the $\{557\}_\gamma$ HPs. *Ex situ* or *in situ* TEM observations of the structure of dislocations at the intersection of two $\{111\}_\gamma$ glide planes could also be performed similarly to the work performed by Spencer et al. [66]

for the formation of martensite at the intersection of two ε plates, but now in order to detect some special dislocation pile-ups and try to identify them as wedge disclinations $\mathbf{w}_A$ and $\mathbf{w}_B$. Particular attention can also be paid to possible formation of Lomer-Cottrel locks. High Resolution TEM (HRTEM) studies on the dislocations at the interphase boundaries are also particularly interesting. The intermediate transient lattice and the sets of extra-half $(\bar{1}11)_\gamma$ planes on few nm at the austenite and martensite interface observed by Ogawa and Kajiwara [67][68] on HRTEM images along the electron beam direction x = $[111]_\alpha$ = $[110]_\gamma$ are interesting features that seem in agreement with the disclination $\mathbf{w}_A$. The two sets of screw dislocations with Burgers vectors $b_1$ = ½ $[111]_\alpha$ = ½ $[110]_\gamma$ and $b_2$ = ½ $[\bar{1}11]_\alpha$ shown by Shibata et *al.* [69] on HRTEM images along the electron beam $[\bar{1}10]_\alpha$ = $[001]_\gamma$ could correspond to the disclination $\mathbf{w}_{B.}$ It could be interesting to revisit these observations in the light of the one-step theory. We have also hopes in the ultra-fast synchrotron X-ray diffraction experiments we have acquired recently [33]. Of course, as initially thought, the data will be treated in order to try to detect the fugitive intermediate hcp phase imagined in our first two-step model; however, since we are now very confident in the new one-step model, we will also treat them to find some elements of confirmations of the Pitsch distortion by according a particular attention to the neutral line.

Simulations seem also to be an interesting tool that could be used to find favourable energetic arguments to the Pitsch versus Bain distortion, or even versus the two-step KSN model. For example, it was shown recently by Sandoval *et al.* [70] by atomistic simulations that the Bain path implies compressive pressures five order higher than the KSN path. We were also impressed by the results obtained recently by Sinclair *et al.* [71][72] showing by molecular dynamics computed on the base of the Boger and Burgers mechanism and Olson and Cohen model [25][26][27], that Pitsch OR is obtained at the intersection of faulted bands. They also added that "*the rapid increase in growth kinetics corresponds to a change in orientation relationship from Pitsch to Kurdjumov-Sachs*", which is in agreement with our model of reorientation of the Pitsch variants in the strain field of martensite transformation (section 3.2). It could be very interesting to perform simulations as Sandoval and Sinclair did, but now on the base of the Pitsch distortion and associated one-step model.

Since the continuous patterns in the EBSD or X-ray diffraction pole figures were also observed in bainitic steel and some brass alloys, we believe that the one-step theory also applies for these materials.

# 5 Conclusions

We proposed a new theory of $\gamma \rightarrow \alpha$ martensitic transformation. This theory is born from the conviction that the continuous rotations A and B observed in EBSD pole figures of the martensite variants inside prior austenitic grains can be interpreted as trace of the transformation mechanisms. Our first attempt to build a theory based on this idea lead us to propose a two-step model implying an intermediate hcp phase. In that theory rotations A and B were each consequence of a distinct step: rotation A from the fcc-hcp step and rotation B from hcp-bcc step. By a careful and rigorously justified classification of the crystallographic structures and substructures (packets) of the KS variants, we realized that the rotations A(10.5°) and B(10.5°) are actually two operators between KS variants, the former in the CPD packets and the latter in the CPP packets. Moreover the rotations A(5.25°) and B(5.25°) links the 24 KS variants to the 12 Pitsch and 12 NW variants respectively. The crystallographic intricacy between the CPDs and CPPs of the martensitic variants can therefore be summarized by a continuous path between the KS, NW and Pitsch variants. We came to the conviction that the A and B continuous rotations are in fact the consequence of a unique mechanism. One of the three ORs (KS, NW, Pitsch) should be sufficient to precise that mechanism and the two other ones should be the consequence of it. We studied in detail the Pitsch OR and the Pitsch distortion, and built a simple and self consistent model of martensitic transformation.

The Pitsch distortion respects the hard sphere packing of the iron atoms. It consists in (a) a 19.5° distortion of the $(001)_\gamma$ plane into the $(\bar{1}10)_\alpha$ plane while keeping dense the CPDs $[\bar{1}10]_\gamma$, $[1\bar{1}0]_\gamma$ and $[111]_\alpha$, $[11\bar{1}]_\alpha$, and (b) a 10.5° distortion of the $(110)_\gamma$ plane into the $(111)_\alpha$ plane. There is a neutral line along the close packed direction: $[110]_\gamma$ // $[111]_\alpha$. No shuffle is required for the transformation. The Pistch distortion can be expressed by a simple diagonal matrix (in a non orthogonal reference frame) with principal strains of 0%, -5.8% and 15.5%, therefore well below the +12%, +12%, -20% values of the Bain distortion (which was thought to be impossible because Bain was assumed in the past to be the best choice). The calculations based on the Pitsch tensor proved that the $(\bar{1}11)_\gamma$ plane is also transformed into the $(110)_\alpha$ plane. The rotations A and B can thus be explained as follows: during the formation of the nucleus by Pitsch distortion, the surrounding austenite matrix is deformed in order to accommodate the 10.5° distortions of the $(110)_\gamma$ and $(\bar{1}11)_\gamma$ planes. During growth, the martensite continues to form by Pitsch distortion in the deformed matrix in Pitsch OR locally, and then gradually becomes misoriented by rotations of 10.5° around the $[110]_\gamma$ and $[\bar{1}11]_\gamma$ directions, i.e. by the A and B rotations respectively, to finally reach KS and NW OR in reference to the matrix far from the transformation. The rotations A and B, and the KS and NW ORs, are all the consequence of the unique one-step mechanism, i.e. the Pitsch distortion and a gradual reorientation of the variants during their growth in the deformation field of the surrounding matrix created by the transformation itself.

This theory is the first simple physical theory of martensitic transformation. It has many advantages in comparison to the usual PTMT; it is based on a distortion with strains lower than Bain, it leads directly, without any complex matrix calculations or dubious assumptions, to an OR and an internal gradient of ORs observed experimentally. Many microstructural features reported in literature can be re-interpreted with it, at the price of a change of paradigm. Most of the time, the elongated direction of the martensite, whatever its shape (needles, lath or lenticular plates), is or contains the neutral line along $[110]_{\gamma}$ direction. The $\{225\}_{\gamma}$ habit planes do not correspond to an "invariant plane strain" as in PTMT, but simply to low index $\{112\}_{\alpha}$ facets of the martensite nucleus. The "twins" sometimes observed at the midrib are not the result of an "inhomogeneous lattice invariant shear" but are actually Pitsch variants created by the phase transformation itself. The effect of prior plastic deformation of austenite can be explained by the creation of intersection points between $\{111\}_{\gamma}$ glide planes, probably Lomer-Cottrel locks, generating dislocation pile-ups and disclinations along the neutral line $<110>_{\gamma}$ with a distortion field that triggers the Pitsch distortion and therefore the martensitic transformation during cooling. Many results obtained in the past can now be revisited in the light of the one-step theory.

## ACKNOWLEDGMENTS

We would like to thank the persons I discussed with about EBSD and martensite, Stephan Zaefferer, Goro Myamoto, Hisashi Sato, Yves Bréchet, Muriel Véron, Laure Guétaz, Pasi Suikkanen, Yann de Carlan, Eric Payton, and the persons who got involved in ultra-fast synchrotron X-ray diffraction, Andrew King, Dominique Thiaudière, Louis Hennet and Jean-Luc Béchade. I hope in the future to extract as much as possible of all the data we have acquired together.